\documentclass[12pt,a4paper,twoside]{article}

\usepackage{amsmath}
\usepackage{amssymb}

\numberwithin{equation}{section} 
\newtheorem{theorem}{Theorem}[section]
\newtheorem{lemma}[theorem]{Lemma}
\newtheorem{definition}[theorem]{Definition}
\newtheorem{example}[theorem]{Example}
\newtheorem{proposition}[theorem]{Proposition}
\newtheorem{corollary}[theorem]{Corollary}
\newenvironment{proof}{\paragraph{Proof.}}{\hfill $\square$\\}
\newenvironment{proof*}{\paragraph{Proof.}}{}

\newcommand{\arrow}{\rightarrow}
\newcommand{\map}{\mapsto}
\newcommand{\bb}[1]{\mathbb{#1}}
\newcommand{\Diff}[3]{\left . \frac{d}{d#1}#2\right |_{#3}}
\newcommand{\kr}[2]{\left . #1\right |_{#2}}
\newcommand{\alg}{\mathfrak{g}}
\newcommand{\Alg}{\mathcal{A}}
\newcommand{\Si}{\mathbb{S}^1}
\newcommand{\Cm}{\mathbb{C}}
\newcommand{\smf}{\mathcal{C}^\infty}
\newcommand{\Rm}{\mathbb{R}}
\newcommand{\pr}{\partial}
\newcommand{\me}{\geqslant}
\newcommand{\les}{\leqslant}
\newcommand{\Dx}[1]{\partial_{x}^{#1}}
\newcommand{\bra}[1]{\left (#1\right )}
\newcommand{\brac}[1]{\left [#1\right ]}
\newcommand{\pobr}[1]{\left \{#1\right \}}
\newcommand{\Tr}{{\rm Tr}}
\newcommand{\var}[2]{\frac{\delta #1}{\delta #2}}
\newcommand{\ad}{{\rm ad}}
\newcommand{\eqreff}[2]{(\ref{#1}-\ref{#2})}
\newcommand{\res}{{\rm res}}
\newcommand{\pmatrx}[1]{\begin{pmatrix} #1 \end{pmatrix}}

\begin{document}

\markboth{B.M. Szablikowski  and M. B\l aszak}{Meromorphic Lax
representations of dispersionless systems}

\title{Meromorphic Lax representations of (1+1)-dimensional multi-Hamiltonian
dispersionless systems}

\author{B\l a\.zej M. Szablikowski\footnote{E-mail: bszablik@amu.edu.pl }$\ $
and Maciej B\l aszak\footnote{E-mail: blaszakm@amu.edu.pl}\\
Institute of Physics, A. Mickiewicz University,\\ Umultowska 85,
61-614 Pozna\'n, Poland}



\maketitle

\begin{abstract}
Rational Lax hierarchies introduced by Krichever are generalized.
A systematic construction of infinite multi-Hamiltonian
hierarchies and related conserved quantities is presented. The
method is based on the classical $R$-matrix approach applied to
Poisson algebras. A proof, that Poisson operators constructed near
different points of Laurent expansion of Lax functions are equal,
is given. All results are illustrated by several examples.
\end{abstract}

\section{Introduction}

First order PDE's of the form
\begin{equation*}
(u_i)_t = \sum_{j} A^j_i(u) (u_j)_x\qquad i,j=1,...,n
\end{equation*}
are called hydrodynamic or dispersionless systems in
(1+1)-dimension. An important subclass of such systems are these
which have multi-Hamiltonian structure, infinite hierarchy of
symmetries and conservation laws. Differential Poisson structures
for hydrodynamic systems were introduced for the first time by
Dubrovin and Novikov \cite{Du1} in the form \eqref{novdub} with
$c=0$, where $g^{ij}$ is a contravariant nondegenerate flat metric
and $\Gamma _{k}^{ij}$ are related coefficients of the
contravariant Levi-Civita connection. Then, they were generalized
by Mokhov and Ferapontov \cite{Fe} to the nonlocal form
\begin{equation}\label{novdub}
\pi _{ij}=g^{ij}(u)\partial _{x}-\sum_{k}\Gamma _{k}^{ij}(u)
(u_k)_x + c  (u_i)_x \Dx{-1} (u_j)_x
\end{equation}
in the case when $g^{ij}$ is of constant curvature $c$. The
natural geometric setting of related bi-Hamiltonian structures
(Poisson pencils) is the theory of Frobenious manifolds based on
the geometry of pencils of contravariant metrics \cite{Du2}.
Nevertheless, the condition of nondegeneracy of $g^{ij}$ for the
above Poisson tensors is not necessary. The degenerate
hydrodynamic Poisson tensors were considered by Grinberg \cite{G}
and Dorfmann \cite{D}.

In paper \cite{K} Krichever introduced integrable dispersionless
systems with rational Lax functions on $\Cm P^1$ of the form
\begin{equation}\label{kri}
L = p^N + \sum_{k=0}^{N-1}a_k p^k +
\sum_{l=1}^{\alpha}\sum_{i=1}^{i_l}\frac{a_{l,i}}{(p -
p_l)^i}\qquad n\me 0,\quad i_l\me 0,
\end{equation}
where $a$'s and the poles $p_l$ are smooth dynamical fields. Then,
around all poles of \eqref{kri}, i.e. $\infty$ and $p_l$, the
powers of Laurent expansions of $L$ generate infinite Lax
hierarchies of commuting vector fields with Lie bracket being the
canonical Poisson bracket \eqref{pb} with $r=0$. Moreover, near
these poles one can construct infinite hierarchies of constants of
motion. Rational Lax functions \eqref{kri} with related Lax
hierarchies have been introduced in \cite{K} in the context of
Whitham hierarchies and topological field theories. From this
point of view they have been considered also in \cite{AK} and
\cite{AK2}. The bi-Hamiltonian structures of Benney and Toda like
Lax hierarchies, but with Poisson bracket \eqref{pb} with $r=1$
and rational Lax functions was developed in \cite{FS}. Their
various reductions were also studied. They also have been
investigated in the context of degenerate Frobenius manifolds
\cite{S}. In \cite{Z}, it was shown how to construct recursion
operators for some classes of such rational Lax representations.

In the theory of nonlinear evolutionary PDE's (dynamical systems)
one of the most important problems is a systematic construction of
integrable systems. By integrable systems we understand those
which have infinite hierarchy of commuting symmetries. It is well
known that a very powerful tool, called the classical $R$-matrix
formalism, can be used for systematic construction of
(1+1)-dimensional field and lattice integrable dispersive systems
(soliton systems) \cite{STS}-\cite{BS} as well as dispersionless
integrable field systems \cite{Li}-\cite{BS2}. Moreover, the
$R$-matrix approach allows a construction of Hamiltonian
structures and conserved quantities.

In this paper the systematic approach of classical $R$-matrices to
(1+1)-integrable dispersionless multi-Hamiltonian systems with
meromorphic Lax hierarchies is presented. In the frames of that
formalism we generalize the results of Krichever onto a wider set
of integrable hierarchies with rational Lax representations as
well as we develop systematically their multi-Hamiltonian
structures. Section 2 briefly presents a number of basic facts and
definitions concerning the formalism of $R$-matrices on Poisson
algebras. In section 3 we define Poisson algebras of meromorphic
functions and construct $R$-matrices. We study multi-Hamiltonian
structures and show the main theorem, that Poisson tensors
constructed for fixed Poisson algebra at different points of
Laurent expansions of $L$ are equal and that related hierarchies
mutually commute. In section 4 we investigate appropriate forms of
meromorphic Lax functions, with finite number of dynamical fields,
which permit construction of integrable dispersionless systems and
illustrate results by a large number of examples.

\section{Classical $R$-matrix theory on Poisson algebras}

The crucial point of the formalism is the observation that
integrable dynamics from some functions space can be represented
by integrable dynamics from an appropriate Lie algebra in the form
of Lax equation
\begin{equation}\label{laxdyn}
  L_t = \ad_A^* L = \brac{A,L},
\end{equation}
i.e. a coadjoint action of some Lie algebra $\alg$ on its dual
$\alg^*$, with the Lax operators $L$ taking values from this Lie
algebra $\alg^* \cong \alg$, where $[\cdot, \cdot]$ is an
appropriate Lie bracket. From \eqref{laxdyn} it is clear that we
confine to such algebras $\alg$ for which its dual $\alg^*$ can be
identified with $\alg$ through the duality map $\langle
\cdot,\cdot \rangle: \alg^* \times \alg \arrow \bb{R}$. So, we
assume the existence of a scalar product $(\cdot, \cdot)$ on
$\alg$ which is symmetric, non-degenerate and $\ad$-invariant:
$(\ad_a b,c)_\alg + (b,\ad_a c)_\alg = 0$. This abstract
representation \eqref{laxdyn} of integrable systems is referred to
as the Lax dynamics. Obviously, we have one-to-one correspondence
between given lax dynamics and original dynamics.

On the space of smooth functions on the dual algebra $\alg^*$
there exists a natural Lie-Poisson bracket
\begin{equation}\label{liepo}
\pobr{H,F}(L):=\langle L, \brac{dF,dH} \rangle \qquad L\in \alg^*
\quad H, F\in \smf \bra{\alg^*},
\end{equation}
where $dF$, $dH$ are differentials belonging to $\alg$ which can
be calculated from
\begin{equation}\label{grad}
\Diff{t}{F(L+tL')}{t=0} = \langle L',dF(L) \rangle,\qquad L,L'\in
\alg^*.
\end{equation}

A linear map $R:\alg \arrow \alg$, such that the bracket
\begin{equation}\label{rbra}
  \brac{a,b}_R := \brac{R a, b} + \brac{a,R b}
\end{equation}
is a second Lie product on $\alg$ is called the classical
$R$-matrix. We will additionally assume that $R$-matrices commute
with derivatives with respect to evolution parameters, i.e.
\begin{equation}\label{assum}
    \bra{R L}_t = R L_t .
\end{equation}
This property is equivalent to the assumption that $R$ commutes
with differentials of smooth maps from $\alg$ to $\alg$. This
property is used in the proof of Theorem~4.2 in \cite{Li},
although not explicitly stressed there. The equality \eqref{assum}
will be used in subsection 3.5 to show a commutation between
particular Lax hierarchies.

\begin{definition}
Let $\Alg$ be a commutative, associative algebra with unit. If
there is a Lie bracket on $\Alg$ such that for each element $a\in
\Alg$, the operator $\ad_a:b\map \{a,b\}$ is a derivation of the
multiplication, i.e. $\{a,bc\} = \{a,b\}c+b\{a,c\}$, then
$(\Alg,\{\cdot,\cdot\})$ is called a Poisson algebra and bracket
$\pobr{\cdot,\cdot}$ is a Poisson bracket.
\end{definition}
Thus, the Poisson algebras are Lie algebras with an additional
structure.
\begin{theorem} \cite{Li}
Let $\Alg$ be a Poisson algebra with Poisson bracket
$\{\cdot,\cdot\}$ and non-degenerate $\ad$-invariant scalar
product $(\cdot,\cdot)$ with respect to which the operation of
multiplication is symmetric, i.e. $(ab,c)=(a,bc)$, $\forall
a,b,c\in \Alg$. Assume $R$ is a classical $R$-matrix, such that
\eqref{assum} holds, then for each integer $n\me 0$, the formula
\begin{equation}\label{pobr}
\pobr{H,F}_n = \bra{L, \pobr{R(L^ndF),dH}+\pobr{dF,R(L^ndH)}}
\end{equation}
where $H,F$ are smooth functions on $\Alg$, defines a Poisson
structure on $\Alg$. Moreover, all $\{\cdot,\cdot\}_n$ are
compatible.
\end{theorem}
The related Poisson bivectors $\pi_n$, such that $\{H,F\}_n =
(dF,\pi_ndH)$ are given by the following Poisson maps
\begin{equation}\label{pot}
\pi_n :dH\map \pobr{R(L^ndH),L} + L^n R^*\bra{\pobr{dH,L}},\qquad
n\me 0
\end{equation}
where the adjoint of $R$ is defined by the relation
$(R^*a,b)=(a,Rb)$. Notice that the bracket \eqref{pobr} with $n=0$
is just a Lie-Poisson bracket with respect to the Lie bracket
\eqref{rbra}. Referring to the dependence on $L$, Poisson maps
\eqref{pot} are called linear for $n=0$, quadratic for $n=1$ and
cubic for $n=2$, respectively.

We will look for a natural set of functions in involution w.r.t.
the Poisson brackets \eqref{pobr}. Such functions are Casimir
functions of the natural Lie-Poisson bracket \eqref{liepo}. A
sufficient condition for smooth function $F(L)$ to be a Casimir
function is that its differential $dF\in \ker \ad_L$, i.e.
$[dF,L]=0$. Hence, the following Lemma is valid
\begin{lemma}\cite{Li}\label{lemma}
Smooth functions on $\Alg$ which are Casimir functions of the
natural Lie-Poisson bracket \eqref{liepo} commute with respect to
$\{\cdot,\cdot\}_n$. The Hamiltonian system generated by a Casimir
function $C(L)$ and the Poisson structure $\{\cdot,\cdot\}_n$ is
given by the Lax equation
\begin{equation}\label{laxhier}
L_t = \brac{R(L^ndC),L},\qquad L\in \Alg.
\end{equation}
\end{lemma}

Let us assume that an appropriate scalar product on Poisson
algebra $\Alg$ is given by the trace form $\Tr: \Alg \arrow
\bb{R}$, such that
\begin{equation*}
\bra{a,b} = \Tr \bra{ab}.
\end{equation*}
As we have assumed a nondegenerate trace form $\Tr$ on $\Alg$, we
will consider the most natural Casimir functionals given by the
trace of powers of $L$, i.e.
\begin{equation}\label{casq}
dC_q(L) = L^q \Longleftrightarrow
\begin{cases}
C_q(L) = \frac{1}{q+1} \Tr \bra{L^{q+1}} & \text{for } q\neq -1\\
C_{-1}(L) = \Tr \bra{\ln L} & \text{for } q=-1
\end{cases}
\end{equation}
for which the related gradients follows by \eqref{grad}. Then,
taking these $C_q(L)$ as Hamiltonian functions, one finds a
hierarchy of evolution equations which are multi-Hamiltonian
dynamical systems
\begin{equation}\label{eveq}
L_{t_q} = \pobr{R(dC_q),L} = \pi_{0}(dC_q) = \pi_{1}(dC_{q-1}) =
... = \pi_{l}(dC_{q-l}) = ...\ .
\end{equation}
For any $R$-matrix each two evolution equations in the hierarchy
\eqref{eveq} commute due to the involutivity of the Casimir
functions $C_q$. Each equation admits all the Casimir functions as
a set of conserved quantities in involution. In this sense we will
regard \eqref{eveq} as a hierarchy of integrable evolution
equations.

To construct the simplest $R$-structure let us assume that the
Poisson algebra $\Alg$ can be split into a direct sum of Lie
subalgebras $\Alg_+$ and $\Alg_-$, i.e. $\Alg =\Alg_+ \oplus
\Alg_-$, $[\Alg_\pm,\Alg_\pm]\subset \Alg_\pm$. Denoting the
projections onto these subalgebras by $P_\pm$, the classical
$R$-matrix is well defined as
\begin{equation}\label{rp}
R = \tfrac{1}{2} (P_+ - P_-) = P_+ - \tfrac{1}{2} = \tfrac{1}{2} -
P_-.
\end{equation}

Following the above scheme, we are able to construct in a
systematic way integrable multi-Hamiltonian dispersionless
systems, with infinite hierarchy of involutive constants of motion
and infinite hierarchy of related commuting symmetries, on an
appropriate Poisson algebras. Finally, in the last step, we
reconstruct our multi-Hamiltonian hierarchies in the original
function space of related dispersionless systems.

\section{Lax hierarchies for dispersionless systems}

\subsection{Poisson algebras of meromorphic functions}

Let $\mathcal{F}$ be the algebra of meromorphic functions with a
finite number of poles, i.e. these analytic functions which have
no essential singularities, on a Riemann sphere $\Cm P^1$ (i.e.
complex plane with point at $\infty$). Let $p$ be a point in $\Cm
P^1$. Assume now that this algebra depends effectively on an
additional spatial variable $x\in \Omega$. Denote by $\Alg$ the
algebra of all smooth functions: $f: \Omega \rightarrow
\mathcal{F}$, i.e. $\Alg = \smf(\Omega,\mathcal{F})$. Let $\Omega
= \Si$ if we assume these functions to be periodic in $x$ or
$\Omega = \Rm$ if these functions supposed to belong to the
Schwartz space for a fixed parameter $p$. The Poisson bracket on
$\Alg$ can be introduced in infinitely many ways as
\begin{equation}\label{pb}
\pobr{f, g}_r := p^r \bra{\pr_pf\pr_xg-\pr_xf\pr_pg}\qquad r\in
\bb{Z}\qquad f,g\in \Alg.
\end{equation}
Then, fixing $r$, $\Alg$ is the Poisson algebra with an
appropriate bracket \eqref{pb}. Poisson brackets \eqref{pb} are
generalization of canonical Poisson bracket ($r=0$) through the
addition of $p^r$ factor.

To construct classical $R$-matrices we have to decompose $\Alg$
into a direct sum of Lie subalgebras. It can be done by expanding
functions belonging to $\Alg$ in an appropriate annulus near a
given point $\lambda$. Three kind of points on parametrized by $x$
$\Cm P^1$ will be important. Two fixed points: $\infty$ and $0$,
as well as points being smooth fields $v(x)$ from $\Omega$ to $\Cm
P^1$.

\subsection{Classical $R$-matrices}

Once we fixed Poisson algebra we are able to construct
$R$-matrices and related Lax vector fields for which the algebra
$\Alg$ constitutes the phase space.

\paragraph{The expansion around $\infty$.}

First, let us consider the case of point at $\infty$. Then,
meromorphic functions from $\Alg$ expanded around $\infty$ are
given by Laurent series:
\begin{equation}\label{ai}
\Alg^\infty = \pobr{\sum^N_{i= -\infty} a_i(x)p^i },
\end{equation}
where $a_i(x)$ are dynamical fields. To construct $R$-matrices we
have to decompose $\Alg_\infty$ into Lie subalgebras. For a fixed
$r$ let $\Alg^\infty_{\me k-r} = \{\sum_{i\me k-r} a_i(x)p^i\}$
and $\Alg^\infty_{<k-r} = \{\sum_{i< k-r} a_i(x)p^i\}$. Let $a
p^m$ and $b p^n$ be elements from \eqref{ai} of order $m$ and $n$,
respectively. Poisson bracket \eqref{pb} between these elements
has the order $m+n+r-1$ as
\begin{equation*}
    \pobr{ap^m, bp^n}_r = \bra{m ab_x-n a_xb}p^{m+n+r-1}.
\end{equation*}
Now, simple inspection shows that $\Alg^\infty_{\me k-r}$ and
$\Alg^\infty_{< k-r}$ are Lie subalgebras in the following cases:
\begin{enumerate}
    \item $r=0$, $k=0$;
    \item $r\in \bb{Z}$, $k=1,2$;
    \item $r=2$, $k=3$.
\end{enumerate}
So, fixing $r$ we fix the Lie algebra structure with $k$ numbering
the $R$-matrices \eqref{rp} given in the following form
\begin{equation}\label{ri}
R = P^\infty_{\me  k-r} - \tfrac{1}{2}
\end{equation}
where $P^\infty_{\me  k-r}$ is an appropriate projection onto Lie
subalgebra of functions expanded in Laurent series \eqref{ai}. So,
Lax hierarchy \eqref{laxhier} assigned by \eqref{ri} for a given
Lax function $L\in \Alg^{\infty}$ is
\begin{equation}\label{laxi}
    L_{t_q} = \pobr{\bra{L^{\tfrac{q}{N}}}^\infty_{\me  k-r},L}_r\qquad q\in \bb{Z}_+,
\end{equation}
where $(\cdot)^\infty_{\me  k-r}\equiv P^\infty_{\me k-r}(\cdot)$
and $N\neq 0$ is the highest order of $L$ expanded in Laurent
series at $\infty$. So, if $L$ has pole at $\infty$ then $N>0$ and
the powers are positive, or if $L^{-1}$ has pole at $\infty$ then
$N<0$ and the powers are negative.

\paragraph{The expansion around $0$.}

Meromorphic functions expanded near $0$ constitute the following
algebra
\begin{equation*}\label{a0}
\Alg^0 = \pobr{\sum^{\infty}_{i= -m} a_i(x)p^i }.
\end{equation*}
The situation here is similar to the previous case. So,
$R$-matrices are defined for the same $r$ and $k$ as at $\infty$
and are of the form
\begin{equation*}\label{r0}
R = \frac{1}{2} - P^0_{<  k-r}.
\end{equation*}
Hence, Lax hierarchies are
\begin{equation}\label{lax0}
    L_{t_q} =  - \pobr{\bra{L^{\tfrac{q}{m}}}^0_{<  k-r},L}_r\qquad q\in \bb{Z}_+,
\end{equation}
where $-m\neq 0$ is the lowest order of Laurent series of $L$
expanded around $0$. So, if $L$ has pole at $0$ then $m>0$ and the
powers are positive, while if $L^{-1}$ has pole at $0$ then $m<0$
and the powers are negative.

Now, we will show that schemes for points at $\infty$ and at $0$
are interrelated.
\begin{proposition}\label{0i}
Under the transformation
\begin{equation*}\label{tran}
  x'=x\quad p'=p^{-1}\quad t'=t
\end{equation*}
the Lax hierarchy \eqref{laxi} defined by $L\in \Alg^\infty$ for
$r,k$ transforms into the Lax hierarchy \eqref{lax0} defined by
 $L'=L\in \Alg^0$ for $k'=3-k,r'=2-r$, i.e.
\begin{equation*}\label{trans}
L \text{ for } k,\ r \text{ at } \infty \quad \Longleftrightarrow
\quad L'=L \text{ for } k'=3-k,\ r'=2-r \text{ at } 0.
\end{equation*}
\end{proposition}
\begin{proof}
It follows from the observation that $\{\cdot,\cdot\}_r =
p^r\pr_p\wedge\pr_x = -p'^{2-r}\pr_{p'}\wedge\pr_{x'} = -
\{\cdot,\cdot\}'_{r'}$ and $(L^n)^\infty_{\me  k-r} =
({L'}^n)^0_{\les  k-r} = ({L'}^n)^0_{< k'-r'}$.
\end{proof}

\paragraph{The expansion around $v(x)$.}

Now, we will consider meromorphic functions in the form of Laurent
series expanded around some field $v(x)$:
\begin{equation*}\label{av}
\Alg^v = \pobr{\sum^{\infty}_{i= -m} a_i(x)(p-v(x))^i }.
\end{equation*}
Notice, that $v(x)$ is a dynamical field of the same kind as
coefficients $a_i(x)$. Let $\Alg^v_{\me k-r} = \{\sum_{i\me k-r}
a_i(p-v)^i\}$ and $\Alg^v_{< k-r} = \{\sum_{i< k-r} a_i(p-v)^i\}$.
Here the situation is a bit more complicated as one has to expand
$p^r$ in \eqref{pb} at $v(x)$, i.e.
\begin{equation*}
p^r = \sum_{s=0}^\infty \tbinom{r}{s}v(x)^{r-s}(p-v(x))^s
\end{equation*}
where $\binom{r}{s} = (-1)^s \binom{-r+s-1}{s}$ for $r<0$. Hence,
$p^r$ as the element of $\Alg_v$, has the lowest order equal zero,
the highest order equal $r$ for $r\me 0$ and infinity for $r<0$.
Therefore
\begin{equation*}\label{pbv}
    \pobr{a(p-v)^m, b(p-v)^n}_r = \bra{(p-v)^\alpha+ ... + v^r}
    \times \bra{mab_x-na_xb}(p-v)^{m+n-1},
\end{equation*}
where $\alpha=r$ for $r\me 0$ and $\alpha$ goes to $\infty$ for
$r<0$. One finds that $\Alg^v_{\me  k-r}, \Alg^v_{<  k-r}$ are Lie
subalgebras in the following cases:
\begin{enumerate}
\item $r=0$, $k=0,1,2$; \item $r=1$, $k=1,2$; \item $r=2$, $k=2,3$
\end{enumerate}
and $R$-matrices have the form
\begin{equation}\label{rv}
R =  \frac{1}{2} - P^v_{<k-r} .
\end{equation}
However, we have to choose these $R$-matrices which commutes with
derivatives with respect to evolution parameters. Let $L= \sum_i
a_i (p-v)^i$. Then,
\begin{align*}
(RL)_t - R L_t &= P^v_{<k-r}L_t - \bra{P^v_{<k-r}L}_t\\
&= P^v_{<  k-r}\bra{\sum_i (a_i)_t (p-v)^i - \sum_i i a_i v_t
(p-v)^{i-1}} - \frac{d}{dt}\bra{\sum_{i<k-r} a_i (p-v)^i}\\
&= (k-r)a_{k-r-1} v_t (p-v)^{k-r-1}
\end{align*}
and equality \eqref{assum} holds when $k-r=0$. Hence, further on
we will consider only $R$-matrices \eqref{rv} for
\begin{equation*}
k=r=0,1,2.
\end{equation*}
In consequence, one finds the following Lax hierarchies related to
$R$-matrices \eqref{rv}
\begin{equation*}\label{laxv}
L_{t_q} =  - \pobr{\bra{L^{\tfrac{q}{m}}}^v_{< 0},L}_r\qquad q\in
\bb{Z}_+\quad r=0,1,2,
\end{equation*}
where $-m\neq 0$ is the lowest order of Laurent series of $L$ at
$v$.

\subsection{Scalar products}

To construct Poisson structures one has to define an appropriate
scalar product on $\Alg $. We will define it near a given point
$\lambda$ by means of the trace form in the algebra $\Alg$ with
the Poisson structure \eqref{pb} for fixed $r$:
\begin{align*}
\Tr_\infty f &= -\int_\Omega \res_\infty \bra{p^{-r}f} dx,\\
\Tr_\lambda f &= \int_\Omega \res_\lambda \bra{p^{-r}f} dx,\qquad
\qquad \lambda = 0,v(x)\quad f\in \Alg,
\end{align*}
where $\res$ is the standard residue. In further considerations
the residue theorem will be very useful. Let $f\in \Alg$ and
$\Gamma$ be a set of all finite poles of $f$. Then, according to
the residue theorem
\begin{equation}\label{rest}
    \sum_{i\in  \Gamma} \res_{\lambda_i}f =
    \frac{1}{2\pi i} \oint_{\gamma_\Gamma} f\ dp \equiv -\res_\infty
    f \qquad \lambda_i \neq \infty
\end{equation}
where $\gamma_\Gamma$ is closed curve encircling all finite poles
of $f$. So, residue at $\infty$ may be different then zero even if
$f$ does not have singularity at this point.
\begin{lemma}
For two arbitrary functions $f,g\in \Alg$ the scalar product:
\begin{equation}\label{prod}
\bra{f,g}_\lambda:=\Tr_\lambda (fg)\qquad \lambda = \infty,0,v(x)
\end{equation}
is symmetric, nondegenerate and $\ad$-invariant.
\end{lemma}
\begin{proof}
The nondegeneracy and symmetry of \eqref{prod} are obvious. Let
$\gamma_\lambda$ be a closed curve circling once a finite pole
$\lambda$, then
\begin{align*}
\Tr_\lambda \pobr{f,g}_r &= \int_\Omega \res_\lambda \bra{\pr_p f \pr_x g} dx - \int_\Omega \res_\lambda \bra{\pr_x f \pr_p g} dx\\
& = \frac{1}{2\pi i}\int_\Omega \oint_{\gamma_\lambda} \bra{\pr_p f \pr_x g} dpdx - \frac{1}{2\pi i}\int_\Omega \oint_{\gamma_\lambda} \bra{\pr_x f \pr_p g} dpdx\\
& = \frac{1}{2\pi i}\int_\Omega \oint_{\gamma_\lambda} \bra{\pr_x
f \pr_p g} dpdx - \frac{1}{2\pi i}\int_\Omega
\oint_{\gamma_\lambda} \bra{\pr_x f \pr_p g} dpdx = 0,
\end{align*}
where we have integrated by parts with respect to $p$ and $x$.
Similar proof is for $\lambda=\infty$. Therefore
\begin{align*}
&\bra{\pobr{f,g}_r,h}_\lambda - \bra{\pobr{g,h}_r,f}_\lambda =
\Tr_\lambda \bra{\pobr{f,g}_r h} -
\Tr_\lambda \bra{\pobr{g,h}_r f} \\
&\qquad \qquad = \Tr_\lambda \bra{\pobr{fh,g}_r-f\pobr{h,g}_r}+
\Tr_\lambda \bra{f\pobr{h,g}_r}= \Tr_\lambda \pobr{fh,g}_r = 0,
\end{align*}
i.e. $\ad$-invariance is proved.
\end{proof}

For a given functional $H(L)\in \smf(\Alg)$ of $L\in \Alg$ the
differential can be calculated by \eqref{grad}. But for functional
$H = \int_\Omega h(u_i) dx $, where $u_i$ are dynamical
coefficient of $L\in \Alg$, we have to show how to construct $dH$.
Differential of $H$ constructed near a given point $\lambda$, will
be denoted by $d_\lambda H\in \Alg$. Coefficients of $d_\lambda H$
depend on dynamical fields and usual variational derivatives
$\var{H}{u_i}$ in such a way that the trace duality assumes the
usual Euclidean form, i.e.
\begin{equation}\label{eucl}
\bra{d_\lambda H,L_t}_\lambda = \Tr_\lambda\bra{d_\lambda H L_t} =
\sum_i \int_\Omega \var{H}{u_i}(u_i)_t\ dx .
\end{equation}
Notice, that from \eqref{eucl} it follows that
\begin{equation}\label{rel}
\forall_{i,j}\quad \bra{d_{\lambda_i}H,K}_{\lambda_i} =
\bra{d_{\lambda_j}H,K}_{\lambda_j}
\end{equation}
where $K$ is vector field on $\Alg$ such that it spans exactly the
same subspace of $\Alg$ as $L_t$.

To find $R^*$, i.e. the adjoint operation to $R$, one has to
determine the adjoint projections near $\lambda$ from the
following relation
\begin{equation*}
\bra{\bra{P^\lambda}^*f,g}_\lambda = \bra{f,P^\lambda
g}_\lambda\qquad f,g\in \Alg^\lambda .
\end{equation*}
So, for $0$ and $\infty$ we have
\begin{equation*}
\bra{P^0_{<  k-r}}^* = 1 - P^0_{< 2r-k}\qquad \bra{P^\infty_{\me
 k-r}}^* = 1 - P^\infty_{\me 2r-k}.
\end{equation*}
The case of $\lambda = v(x)$ is more delicate. Let $A= \sum_m
a_m(p-v)^m$ and $B= \sum_n b_n(p-v)^n$, then for $r\me 0$:
\begin{align*}
\bra{A ,P^v_{<0}B}_v &=  \int_\Omega \res_v \bra{\sum_{s\me
0}\sum_m\sum_{n<0} \tbinom{-r}{s}v^{-r-s}a_mb_n(p-v)^{m+n+s}}dx\\
&= \int_\Omega \sum_{s\me 0}\sum_{n<0}
\tbinom{-r}{s}v^{-r-s}a_{-n-s-1}b_n\ dx = \int_\Omega \sum_{s\me
0}\sum_{m\me -s}
\tbinom{-r}{s}v^{-r-s}a_mb_{-m-s-1}\ dx\\
&= \int_\Omega \res_v \bra{\sum_{s\me 0}\sum_{m\me -s}\sum_n
\tbinom{-r}{s}v^{-r-s}a_mb_n(p-v)^{m+n+s}}dx\\
 &= \bra{p^r \sum_{s\me 0}\tbinom{-r}{s}v^{-r-s}(p-v)^s P^v_{\me -s}A ,
B}_v,
\end{align*}
where we used an appropriate expansion of $p^{-r}$ at $v$. Hence
\begin{equation*}
\bra{P^v_{< 0}}^* = 1 - p^r
\sum_{s=0}^{\infty}\tbinom{-r}{s}v^{-r-s}(p-v)^s P^v_{<-s}\qquad
r\me 0
\end{equation*}
for $r=0$ it reduces to $\bra{P^v_{< 0}}^*  = 1 - P^v_{< 0}$. We
will use simplified notation:
\begin{equation*}
P_v'= \sum_{s=0}^{\infty}\tbinom{-r}{s}v^{-r-s}(p-v)^s P^{v}_{<-s}
\end{equation*}
as then $\bra{P^v_{< 0}}^*  = 1 - p^r P_v'$.

\subsection{Poisson structures}

The Poisson structures \eqref{pobr} at respective points, related
to respective $R$-matrices, are
\begin{equation*}\label{ps}
\pobr{H,F}^n_\lambda = \bra{d_\lambda F,\pi^n_\lambda d_\lambda
H}_\lambda\qquad \lambda = \infty,0,v(x)\quad n\me 0
\end{equation*}
for which Poisson operators are given by the following forms
\begin{align}
\label{poi} &\pi^n_\infty d_\infty H  = \pobr{\bra{L^nd_\infty H}^\infty_{\me k-r},L}_r - L^n\bra{\pobr{d_\infty H,L}_r}^\infty_{\me 2r-k},\\
\notag &\pi^n_0 d_0 H = \pobr{L,\bra{L^nd_0 H}^0_{< k-r}}_r - L^n\bra{\pobr{L,d_0 H}_r}^0_{<2r-k},\\
\notag &\pi^n_v d_v H  = \pobr{L,\bra{L^nd_v H}^v_{<0}}_r -
L^np^rP_v'\bra{\pobr{L,d_v H}_r}.
\end{align}
It is important here to mention that for a given Lax operator $L$
it may happen that $L_t$ does not span a proper subspace of the
full Poisson algebra $\Alg$, i.e. the image of the Poisson
operator $\pi^ndH$ does not coincide with this subspace. Then, in
general, the Dirac reduction can be invoked for restriction of a
given Poisson tensor to a suitable subspace.

\begin{lemma}
The following relations will be needed to prove forthcoming
theorem:
\begin{align*}
& \bra{d_\infty F,\pobr{L,\bra{L^nd_0H}^0_{<k-r}}_r}_\infty =
\bra{d_0H,L^n\bra{\pobr{d_\infty F,L}_r}^\infty_{\me 2r-k}}_0,\\
&\bra{d_\infty F,L^n \bra{\pobr{L,d_0H}_r}_{< 2r-k}}_\infty =
\bra{d_0H,\pobr{\bra{L^nd_\infty F}^\infty_{\me k-r},L}_r}_0,
\end{align*}
for arbitrary $k$ and $r$, and
\begin{align}
\label{rel1} &\bra{d_\infty F,\pobr{L,\bra{L^nd_v
H}^v_{<0}}_r}_\infty =
\bra{d_v H,L^n\bra{\pobr{d_\infty F,L}_r}^\infty_{\me r}}_v,\\
\label{rel2} &\bra{d_\infty F,L^np^rP_v'\bra{\pobr{L,d_v
H}_r}}_\infty = \bra{d_v H,\pobr{\bra{L^nd_\infty F}^\infty_{\me
0},L}_r}_v,
\end{align}
where $r\me 0$.
\end{lemma}
\begin{proof}
We will prove only the first and last relations as for the two
remaining ones the proof is similar. We use property of
$\ad$-invariance and we omit (or add) these elements which do not
contribute in calculations of residues:
\begin{align*}
&\bra{d_\infty F,\pobr{L,\bra{L^nd_0 H}^0_{<k-r}}_r}_\infty = \bra{\bra{L^nd_0 H}^0_{<k-r},\pobr{d_\infty F,L}_r}_\infty \\
&= \int_\Omega \res_\infty \bra{p^{-r}\bra{L^nd_0H}^0_{<k-r}\pobr{L,d_\infty F}_r}dx =  \int_\Omega \res_\infty \bra{p^{-r}\bra{L^nd_0 H}^0_{<k-r}\bra{\pobr{L,d_\infty F}_r}^\infty_{\me 2r-k}}dx\\
&\overset{\text{by \eqref{rest}}}{=} \int_\Omega \res_0
\bra{p^{-r}\bra{L^nd_0H}^0_{<k-r}\bra{\pobr{d_\infty
F,L}_r}^\infty_{\me 2r-k}}dx\\ &=\int_\Omega \res_0
\bra{p^{-r}L^nd_0H \bra{\pobr{d_\infty F,L}_r}^\infty_{\me
2r-k}}dx =  \bra{d_0H,L^n\bra{\pobr{d_\infty F,L}_r}^\infty_{\me
2r-k}}_0 .
\end{align*}
Let $r\me 0$. Using proper expansion of $p^{-r}$ at $v$ we have:
\begin{align*}
&\bra{d_\infty F,L^np^rP_v'\bra{\pobr{L,d_v H}_r}}_\infty = -\int_\Omega \res_\infty \bra{d_\infty F L^nP_v'\bra{\pobr{L,d_v H}_r}}dx\\
&= \int_\Omega \res_\infty \bra{\bra{L^nd_\infty F}^\infty_{\me 0} P_v'\bra{\pobr{d_v H,L}_r}}dx \overset{\text{by \eqref{rest}}}{=} \int_\Omega \res_v \bra{\bra{L^nd_\infty F}^\infty_{\me 0} P_v'\bra{\pobr{L,d_v H}_r}}dx\\
&= \int_\Omega \res_v \bra{\bra{L^nd_\infty F}^\infty_{\me 0} \pobr{L,d_v H}_r}dx = \bra{\bra{L^nd_\infty F}^\infty_{\me 0},\pobr{L,d_v H}_r}_v\\
& = \bra{d_v H,\pobr{\bra{L^nd_\infty F}^\infty_{\me 0},L}_r}_v.
\end{align*}
Thus all relations are valid.
\end{proof}

\begin{theorem}\label{main}
Let $L\in \Alg$ be a meromorphic Lax function. Then for all
appropriate $k$ and $r$
\begin{equation*}\label{mainri}
\pobr{H,F}^n_0 = \pobr{H,F}^n_\infty\quad \text{and}\quad \pi^n_0
d_0 H = \pi^n_\infty d_\infty H
\end{equation*}
while for $k=r=0,1,2$
\begin{equation*}
\forall_i\ \pobr{H,F}^n_{v_i} = \pobr{H,F}^n_\infty \quad
\text{and}\quad \pi^n_{v_i} d_{v_i} H = \pi^n_\infty d_\infty H,
\end{equation*}
where $v_i$ are dynamical fields. Therefore, Poisson structures,
from the original function space of related dispersionless
systems, calculated for fixed $r$ and $k$ at different points are
equal.
\end{theorem}
\begin{proof}
We will prove only the second set of relations as for the first
part the proof is similar. Thus,
\begin{align*}
&\pobr{H,F}^n_{v_i} = \bra{d_{v_i} F,\pi^n_{v_i}d_{v_i} H}_{v_i} \overset{\text{by \eqref{rel}}}{=} \bra{d_\infty F,\pi^n_{v_i}d_{v_i} H}_\infty\\
&= \bra{d_\infty F,\pobr{L,\bra{L^nd_{v_i} H}^{v_i}_{<0}}_r - L^np^rP_{v_i}' \bra{\pobr{L,d_{v_i} H}_r}}_\infty\\
&\overset{\text{by \eqreff{rel1}{rel2}}}{=}
\bra{d_{v_i} H,L^n\bra{\pobr{d_\infty F,L}_r}^\infty_{\me r}-\pobr{\bra{L^nd_\infty F}^\infty_{\me 0},L}_r}_v\\
&= -\bra{d_{v_i} H,\pi^n_\infty d_\infty F}_{v_i,r}
\overset{\text{by \eqref{rel}}}{=} -\bra{d_\infty H,\pi^n_\infty
d_\infty F}_\infty = \pobr{H,F}^n_\infty.
\end{align*}
Now, from the equality of above Poisson brackets it follows that
\begin{align*}
\bra{d_\infty F,\pi^n_\infty d_\infty H}_\infty =
\bra{d_{\lambda_i} F,\pi^n_{\lambda_i}d_{\lambda_i} H}_{\lambda_i}
 \overset{\text{by
\eqref{rel}}}{=} \bra{d_\infty F,\pi^n_{\lambda_i}d_{\lambda_i}
H}_\infty \Longleftrightarrow \pi^n_{\lambda_i} d_{\lambda_i} H =
\pi^n_\infty d_\infty H,
\end{align*}
where $\lambda_i=0,v_i$. Hence the theorem is proved.
\end{proof}

\subsection{Commuting multi-Hamiltonian Lax hierarchies}

Let $L\in \Alg$ be a Lax function such that $L$ and $L^{-1}$ can
have poles at $\infty, 0$ and $v_i(x)$. Then, for appropriate $r$
and $k$ near these poles one can construct the following
multi-Hamiltonian Lax hierarchies \eqref{eveq}
\begin{align}
\label{lhi} L_{t_q} &= \pobr{\bra{L^{\tfrac{q}{N}}}^\infty_{\me
k-r},L}_r = \pi^0_\infty d_\infty H_q^\infty = \pi^1_\infty
d_\infty H_{q-1}^\infty = ...,\\
\label{lh0} L_{\tau_q} &=  - \pobr{\bra{L^{\tfrac{q}{m_0}}}^0_{<
k-r},L}_r =
\pi^0_0 d_0 H_q^0 = \pi^1_0 d_0 H_{q-1}^0 = ...,\\
\label{lhv} L_{\xi_q} &=  -
\pobr{\bra{L^{\tfrac{q}{m_i}}}^{v_i}_{< 0},L}_r = \pi^0_{v_i}
d_{v_i} H_q^{v_i} = \pi^1_{v_i} d_{v_i} H_{q-1}^{v_i} = ...,
\end{align}
where integer $q>0$ and $t, \tau, \xi$ are evolution parameters.
The Hamiltonians are then defined through trace forms near these
poles and are given by \eqref{casq} for $q\me 0$
\begin{equation}\label{ham}
\begin{split}
&H_q^\lambda (L) = \frac{\epsilon}{\frac{q}{n}+1}
\int_\Omega \res_\lambda \bra{p^{-r}L^{\frac{q}{n}+1}}\qquad \text{for } q\neq -n\\
&H_{-n}^\lambda (L) = \epsilon \int_\Omega \res_\lambda
\bra{p^{-r}\ln L}\qquad \text{for } q= -n,
\end{split}
\end{equation}
where $\epsilon = -1, n=N$ for $\lambda=\infty$ and $\epsilon =
1,n=m_0,m_i$ for $\lambda=0,v_i$, respectively. Calculations of
$H_{-n}^\lambda$ from \eqref{ham} for $\lambda$ being the root of
$L$ may cause difficulties as then $\ln L$ has at $\lambda$
essential singularity. There is an alternative approach. First we
look for coefficients of $dH_{-n}^\lambda$ which can be simply
obtained from $\Tr_\lambda (L^{-1} L_t) = \sum_i \int_\Omega
\var{H_{-n}^\lambda}{u_i}(u_i)_t\ dx$, since $d_\lambda
H_{-n}^\lambda = L^{-1}$. Then, we calculate the functional
$H_{-n}^\lambda$ integrating a respective system of equations.

Let us show that Lax hierarchies \eqreff{lhi}{lhv} for fixed $r$
and $k$ mutually commute. Due to Lemma \ref{lemma} Hamiltonians
\eqref{ham}, as Casimirs of the natural Lie-Poisson bracket, are
in involution with respect to Poisson brackets \eqref{ps}, i.e.
$\{H_q^{\lambda_i}, H_{q'}^{\lambda_j}\}^n_{\lambda_k}=0$, where
$\lambda_i=\infty,0,v_i$. From Theorem~\ref{main} it follows that
$\pi^n_{\lambda_i} d_{\lambda_i} = \pi^n_{\lambda_j}
d_{\lambda_j}$. Now, hence $\pi d$ is the Lie algebra
homomorphism, from the algebra of smooth functions to the Lie
algebra of vector fields, the commutation between Lax hierarchies
\eqreff{lhi}{lhv} is immediate. For two vector fields $L_{t_q} =
\pi^n_{\lambda_i} d_{\lambda_i} H_q^{\lambda_i}$ and $L_{t'_{q'}}
= \pi^n_{\lambda_j} d_{\lambda_j} H_{q'}^{\lambda_j}$ we have that
\begin{align*}
\brac{L_{t_q} , L_{t'_{q'}}} = \brac{\pi^n_{\lambda_i}
d_{\lambda_i} H_q^{\lambda_i}, \pi^n_{\lambda_j} d_{\lambda_j}
H_{q'}^{\lambda_j}} = \brac{\pi^n_{\lambda_i} d_{\lambda_i}
H_q^{\lambda_i}, \pi^n_{\lambda_i} d_{\lambda_i}
H_{q'}^{\lambda_j}} = \pi^n_{\lambda_i} d_{\lambda_i}
\pobr{H_q^{\lambda_i}, H_{q'}^{\lambda_j}} = 0,
\end{align*}
where $[\cdot,\cdot ]$ is the Lie bracket between vector fields.
However, for these commutations the Hamiltonian property is not
necessary. We will show it for the Lax hierarchies \eqref{lhi} and
\eqref{lhv} with $k=r=0,1,2$ as for the other combinations the
calculations are similar. We will use simplified notation
$X=L^{\tfrac{q}{N}}, Y=L^{\tfrac{q'}{m_i}}$ and
\begin{equation*}
(X)^\infty_{\me 0}=X^\infty_{\me 0}=X-X^\infty_{< 0}\qquad
(Y)^{v_i}_{<0}=Y^v_{< 0}=Y-Y^v_{\me 0}.
\end{equation*}
Then
\begin{align*}
&\bra{L_{t_q}}_{\xi_{q'}} - \bra{L_{\xi_{q'}}}_{t_q}=\\
 &= \pobr{\bra{X^\infty_{\me 0}}_{\xi_{q'}},L}_r +
\pobr{X^\infty_{\me 0},L_{\xi_{q'}}}_r +
\pobr{\bra{Y^v_{< 0}}_{t_q},L}_r + \pobr{Y^v_{< 0},L_{t_q}}_r\\
 &\overset{\text{by \eqref{assum}}}{=} \pobr{\bra{X_{\xi_{q'}}}^\infty_{\me 0},L}_r +
\pobr{X^\infty_{\me 0},L_{\xi_{q'}}}_r +
\pobr{\bra{Y_{t_q}}^v_{< 0},L}_r + \pobr{Y^v_{<
0},L_{t_q}}_r\\
&= -\pobr{\bra{\pobr{Y^v_{< 0},X}_r}^\infty_{\me 0},L}_r -
\pobr{X^\infty_{\me 0},\pobr{Y^v_{< 0},L}_r}_r\\
&\quad +\pobr{\bra{\pobr{X^\infty_{\me 0},Y}_r}^v_{< 0},L}_r +
\pobr{Y^v_{< 0},\pobr{X^\infty_{\me 0},L}_r}_r\\
&=\pobr{\bra{\pobr{X,Y^v_{< 0}}_r}^\infty_{\me 0} +
\bra{\pobr{X^\infty_{\me 0},Y}_r}^v_{< 0}  -
\pobr{X^\infty_{\me 0},Y^v_{< 0}}_r,L}_r = 0
\end{align*}
where we used Jacoby identity and the last equality holds since for $r=0,1,2$
\begin{align*}
\pobr{X^\infty_{\me 0},Y^v_{< 0}}_r &=
\bra{\pobr{X^\infty_{\me 0},Y^v_{< 0}}_r}^\infty_{\me 0} + \bra{\pobr{X^\infty_{\me 0},Y^v_{< 0}}_r}^v_{< 0}\\
&=
\bra{\pobr{X,Y^v_{< 0}}_r}^\infty_{\me 0} + \bra{\pobr{X^\infty_{\me 0},Y}_r}^v_{< 0}.
\end{align*}
We see now that the restriction \eqref{assum} is indeed crucial.
Actually, in the same way one can prove commutations between
symmetries inside these Lax hierarchies.

Combining results from current and previous subsections we obtain
the following corollary.
\begin{corollary}\label{cor}
Let $L$ be a Lax function in $\Alg$ with fixed Poisson bracket
given by $r$ and let us fix an appropriate $k$. Then, around each
pole of $L$ and $L^{-1}$ one finds infinite hierarchy of commuting
multi-Hamiltonian symmetries and infinite hierarchy of constants
of motion. Moreover, vector fields from these different
 hierarchies mutually commute.
\end{corollary}

In further considerations we are interested in extracting closed
systems with finite number of dynamical functions. Therefore, we
will look for meromorphic Lax functions, with finite number of
dynamical coefficients, which allow a construction of consistent
evolution Lax hierarchies. So, in the following section we will
select an appropriate meromorphic Lax functions.

\section{Meromorphic Lax functions}

The meromorphic Lax function $L$ is an appropriate one if the
right-hand sides of Lax hierarchies \eqreff{lhi}{lhv} can be
written in the form of evolutions $L_t$, i.e. left-hand sides.
These Lax hierarchies are generated by positive and negative
powers (in general fractional) of respective expansions near poles
of $L$ and $L^{-1}$. Actually, the appropriate expansions near
$\infty$ and $0$ are for $k=r=0$; $k=1,2$ and $r\in \bb{Z}$;
$k=3,r=2$, while the expansions near $v_i(x)$ takes place for
$k=r=0,1,2$. One finds these poles by looking for roots of
$L^{-1}$ and $L$, respectively. Important is the following. Let
$L$ be an appropriate Lax function with respect to the Lax
hierarchy related to one of poles. Then, it is as well an
appropriate function for hierarchies for all other allowed poles,
for the same $r$ and $k$. It is so, as by Proposition \eqref{main}
the Lax hierarchy related to one pole can be rewritten for another
one. Moreover, for a given $L$ and fixed $r$ and $k$, the Lax
hierarchies, generated near all poles, will mutually commute.

We would like to investigate the general form of meromorphic Lax
functions being appropriate Lax functions, i.e. such which allow a
construction of integrable dispersionless equations. We will
distinguish between three cases: the first one when $L$ is a
finite formal Laurent series at $0$, the second one when $L$ is a
finite formal Laurent series at pole $v(x)$, and finally more
general case of rational functions.

\subsection{Polynomial Lax functions in $p$ and $p^{-1}$.}

Let us consider Lax functions of the form
\begin{equation}\label{pol0}
L =  u_N p^{N} + u_{N-1} p^{N-1} + ... + u_{1-m} p^{1-m} +
u_{-m}p^{-m},
\end{equation}
i.e. formal finite Laurent series at 0. The coefficients $u_i$ are
dynamical fields. For Lax functions \eqref{pol0}, in general, we
can construct powers near $\infty$ and $0$ which will generate
related Lax hierarchies \eqref{lhi} and \eqref{lh0}, respectively.
If $k=r$ powers calculated around roots of $L$ generate additional
Lax hierarchies given by \eqref{lhv}.

From now on, without loos of generality, we will choose all
appearing constants in the form that will simplify all formulae.
\begin{proposition}\label{apol0}
Lax function of the form \eqref{pol0} is an appropriate one in the
following cases:
\begin{enumerate}
    \item $k=0$, $r=0$: $N\me 2$, $u_N=1$, $u_{N-1} = 0$, $m=0$;
    \item $k=1$, $r\in \bb{Z}$: $N\neq 0$, $u_N=1$, $m\neq 0$ for $r=1$;
    \item $k=2$, $r\in \bb{Z}$: $N\neq 0$ for $r=1$, $m\neq 0$, $u_{-m}=1$;
    \item $k=3$, $r=2$: $N=0$, $m\me 2$, $u_{1-m}=0$, $u_{-m}=1$.
\end{enumerate}
\end{proposition}
We will not prove this proposition as it is the standard case
considered in \cite{BS2}.

\begin{proposition}\label{pp2}
Under the transformation $p'=p^{-1}$ Lax hierarchies, from
Proposition \ref{apol0}, generated by powers calculated at
$\infty$ and $0$ for appropriate $r$ and $k$ transforms into Lax
hierarchies for $0$ and $\infty$ with $r'=2-r$ and $k'=3-k$,
respectively.
\end{proposition}
The proof immediately follows from Proposition \ref{0i}. Notice,
that by transformation $p'=p^{-1}$ Lax hierarchies \eqref{lhv} for
$r=k=0,1,2$ defined at roots of $L$ being dynamical fields fall
out from the scheme presented in this article. On the other hand,
for: $k=1, r=0$; $k=2, r=1$; $k=3, r=2$; according to Proposition
\ref{apol0} one can construct Lax hierarchies only at $\infty$ and
$0$. However, by $p'=p^{-1}$ they transform into cases: $k=2,
r=2$; $k=1, r=1$; $k=0, r=0$; respectively, for which one is able
to construct Lax hierarchies \eqref{lhv} related to all poles
(including poles being dynamical fields) of $L'$ and $L'^{-1}$.
Hence, the relevant cases from Proposition \ref{apolv} are:
\begin{itemize}
    \item $k=0$, $r=0$;
    \item $k=1$, $r\in \bb{Z}\backslash \{0\}$;
    \item $k=2$, $r=2$.
\end{itemize}
The remaining cases can be obtained by transformation $p'=p^{-1}$
according to Proposition \ref{pp2}.

To construct Poisson operators we have to choose a point near
which we will perform the calculations. Nevertheless, as follows
from Theorem \ref{main}, the explicit form of Poisson operators in
the original function space is the same for all points. Thus, we
choose the $\infty$ as it is the standard case. Then, as we
assumed the usual Euclidean form \eqref{eucl}, differentials of
functional $H$ are given by
\begin{equation*}
dH\equiv d_\infty H = \sum_{i=-m}^{N+k-2}\var{H}{u_i}p^{r-1-i},
\end{equation*}
where $m=0$ for $k=0$. Still we have to check whether the above
Lax functions span proper subspaces, w.r.t. Poisson operators
\eqref{poi}, of the full Poisson algebras. We will limit ourselves
to linear ($n=0$) and quadratic ($n=1$) Poisson tensors, as
obviously it is enough to define bi-Hamiltonian structures.
Besides, in the all nontrivial cases Lax functions do not span
proper subspaces w.r.t. Poisson tensors for $n\me 2$.

Poisson tensors restricted to finite number of fields are properly
defined if the highest and lowest orders of $\pi^n_\infty dH$ and
$L_t$ will coincide. Simple inspection shows that the highest
order of $\pi^n_\infty dH$ is equal to $\max \{N+k-2,n N+2r-k-1\}$
and the lowest is $0$ for $k=0$ and $\min \{k-1-m,-n m+2r-k\}$ for
$k=1,2$. Hence, in the case $k=0$ the Lax function always span the
proper subspace w.r.t. the linear Poisson tensor, but for $k=1,2$
only in case when $N\me 2r-2k+1\me -m$, otherwise the Dirac
reduction is required. The linear Poisson tensor is of the form
\begin{equation}\label{lin}
\pi^0_\infty dH  = \pobr{\bra{dH}^\infty_{\me k-r},L}_r
-\bra{\pobr{dH,L}_r}^\infty_{\me 2r-k}.
\end{equation}
The reduced linear tensor for $N=-1$ and $k=r=1,2$ is given by
\eqref{lin_II}. For the quadratic Poisson tensors the Dirac
reduction is always necessary. The calculation procedure of Dirac
reduction is explained in \cite{BS} (in a bit different notation).
The reduced quadratic Poisson tensor for $k=r=0,1,2$ is given by
\begin{equation}\label{quad_I}
\bra{\pi^1_{\infty}}^{red}dH  = \pobr{\bra{L dH}^\infty_{\me
0},L}_r - L \bra{\pobr{dH,L}_r}^\infty_{\me r} +
\frac{1}{N}\pobr{L,\pr_x^{-1}\res_\infty \pobr{dH,L}_0}_r,
\end{equation}
and for $k=1,r=0$ and $k=2,r=1$ takes the form
\begin{equation}\label{quad_II}
\bra{\pi^1_{\infty}}^{red}dH  = \pobr{\bra{L dH}^\infty_{\me
1},L}_r - L \bra{\pobr{dH,L}_r}^\infty_{\me r-1} +
\frac{1}{m}\pobr{L,\pr_x^{-1}\res_\infty \pobr{dH,L}_0}_r.
\end{equation}
Both reduced Poisson tensors are always local as $\res_\infty
\{\cdot,\cdot\}_0 = (...)_x$.

In the article, in general, we present examples for simplest Lax
functions, where calculations are not very much complicated. From
the Lax hierarchies considered we exhibit only the first
nontrivial systems.

\begin{example}Two field system: $k=1$, $r\in \bb{Z}$.

Let us consider the Lax function of the form
\begin{equation}\label{l1}
L = p + u + v p^{-1}.
\end{equation}
It has poles at $\infty$ and $0$. Then, for $\infty$ we have
\begin{align}\notag
&L_{t_{2-r}} = \pobr{\bra{L^{2-r}}^\infty_{\me 1-r},L}_r
\Longleftrightarrow \\ \label{l2}
&\pmatrx{u\\ v}_{t_{2-r}} = (2-r) \pmatrx{(1-r)uu_x-v_x\\
-u_xv-(1-r)uv_x} = \pi_0 dH^\infty_{2-r} = \pi^{red}_1
dH^\infty_{1-r},
\end{align}
where $\bra{L^{2-r}}^\infty_{\me 1-r} = p^{2-r} + (2-r)u p^{1-r}$.
When $r=2$ the next equation from the hierarchy is the first
nontrivial one. For $r=1$ this is the well known dispersionless
Toda system. The hierarchy for $0$ is the same as $L$ has only two
poles of the same order and $(L^q)^\infty_{\me 1-r} = L -
(L^q)^0_{< 1-r}$. The roots of $L$ are $\lambda_\pm =
\tfrac{1}{2}(-u\pm \sqrt{u^2-4v})$. Thus, for $r=1$
\begin{equation*}
(L^{-1})^{\lambda_\pm}_{<0} = \frac{1}{1-\frac{4v}{\lambda_\pm}}
(p-\lambda_\pm)^{-1}
\end{equation*}
and one finds the following equations
\begin{align*}
&L_{\xi^\pm_{-1}} = -\pobr{\bra{L^{-1}}^{\lambda_\pm}_{< 0},L}_1
\Longleftrightarrow \\
&\pmatrx{u\\ v}_{\xi^\pm_{-1}} = \frac{\pm 1}{(u^2-4v)^\frac{3}{2}} \pmatrx{2u_xv-uv_x\\
v(2v_x-uu_x)} = \pi_0 dH^{\lambda_\pm}_{-1} = \pi^{red}_1
dH^{\lambda_\pm}_{-2}.
\end{align*}
Of course, for $k=r=1$ all equations mutually commute.

The Lax function \eqref{l1} the defines proper subspace w.r.t. the
linear Poisson tensor \eqref{lin} only for $r=0,1$. In the cases,
the reduced quadratic Poisson tensors are given by \eqref{quad_II}
and \eqref{quad_I}, respectively. Hence, for $r=0$
\begin{align}\label{l3}
\pi_0 = \pmatrx{0 & \pr\\ \pr & 0}\qquad \pi^{red}_1 =
\pmatrx{2\pr & \pr u\\ u\pr & \pr v+v\pr},
\end{align}
and related Hamiltonians are
\begin{equation*}
H_1^\infty = \int_\Omega uv\ dx\qquad H_2^\infty = \int_\Omega
(u^2v+v^2)\ dx.
\end{equation*}
For $r=1$
\begin{align*}
\pi_0 = \pmatrx{0 & \pr v\\ v\pr & 0}\qquad \pi^{red}_1 =
\pmatrx{\pr v+v\pr & u\pr v\\ v\pr u & 2v\pr v}
\end{align*}
and
\begin{align*}
&H_0^\infty =
\int_\Omega uv\ dx\qquad H_1^\infty = \int_\Omega (u^2v+v^2)\ dx\\
&H_{-2}^{\lambda_\pm} = \int_\Omega \frac{\mp 1}{\sqrt{u^2-4v}}\
dx\qquad H_{-1}^{\lambda_\pm} = \pm \int_\Omega \ln
\frac{u+\sqrt{u^2-4v}}{v}\ dx.
\end{align*}
\end{example}

\begin{example} Two field system: $k=2$, $r=2$.

We will consider Lax function of the form
\begin{equation*}
L = v p + u + p^{-1}
\end{equation*}
i.e. function \eqref{l1} transformed by $p\map p^{-1}$. By
Proposition \ref{pp2} the hierarchy for $\infty$ is given by
hierarchy \eqref{l2} for $k=1$, $r=0$ from above example. The
roots of $L$ are $\alpha_\pm = \tfrac{-u\pm \sqrt{u^2-4v}}{2v}$.
Thus, for
\begin{equation*}
(L^{-1})^{\alpha_\pm}_{<0} = \frac{1}{v-\frac{4v^2}{\alpha_\pm^2}}
(p-\alpha_\pm)^{-1}
\end{equation*}
one finds the following equations
\begin{align*}
&L_{\xi^\pm_{-1}} = -\pobr{\bra{L^{-1}}^{\alpha_\pm}_{< 0},L}_2
\Longleftrightarrow \\
&\pmatrx{u\\ v}_{t_{-1}} = \frac{\pm 1}{(u^2-4v)^\frac{3}{2}} \pmatrx{-uu_x+2v_x\\
2u_xv-uv_x} = \pi_0 dH^{\alpha_\pm}_{-1} = \pi^{red}_1
dH^{\alpha_\pm}_{-2}.
\end{align*}
This system by Proposition \ref{pp2} commutes with \eqref{l2} for
$r=0$. Thus, the Poisson tensors are given by \eqref{l3} with
Hamiltonians
\begin{align*}
H_{-2}^{\alpha_\pm} = \int_\Omega \frac{\pm u}{2\sqrt{u^2-4v}}\
dx\qquad H_{-1}^{\alpha_\pm} = \mp \int_\Omega
\frac{1}{2}\bra{u+\sqrt{u^2-4v}}\ dx.
\end{align*}
\end{example}

\subsection{Polynomial Lax functions in $(p-v)$ and $(p-v)^{-1}$}

Let us consider Lax functions which are formal Laurent series
around $v$, with a finite number of dynamical coefficients, of the
form
\begin{equation}\label{polv}
L = u_N (p-v)^N + u_{N-1}(p-v)^{N-1} + ... + u_{1-m}(p-v)^{1-m} +
u_{-m}(p-v)^{-m}\qquad m\neq 0.
\end{equation}
Lax functions \eqref{polv} have poles at $\infty$ and $v$, near
which calculated powers generate, if allowed by $k$ and $r$,
respective Lax hierarchies. Additional powers with related
hierarchies can be constructed around the roots of $L$.

\begin{proposition}\label{apolv}
Lax function of the form \eqref{polv} is an appropriate one in the
following cases:
\begin{enumerate}
    \item $k=r=0$: $u_N=1$, $u_{N-1} = N v$;
    \item $k=1$, $r\in \bb{Z}$: $N\neq 0$, $u_N=1$, $\kr{L}{p=0}=0$ for $r=1$;
    \item $k=2$, $r\in \bb{Z}$: $N\neq 0$ when $r=1$,$\kr{L}{p=0}=0$, $\kr{\tfrac{d}{dp}L}{p=0}=1$;
    \item $k=3$, $r=2$: $N=0$, $\kr{L}{p=0}=0$,
    $\kr{\tfrac{d}{dp}L}{p=0}=1$, $\kr{\tfrac{d^2}{dp^2}L}{p=0}=0$.
\end{enumerate}
Moreover, for the same $k$ and $r$, the respective Lax hierarchies
commute.
\end{proposition}
\begin{proof}
It is enough to consider the Lax hierarchy related to $\infty$.
Function \eqref{polv} will be appropriate Lax function if the
left- and right-hand sides of Lax hierarchy \eqref{lhi} will
coincide and the number of independent equations will be the same
as the number of dynamical coefficients in $L$. The Lax hierarchy
\eqref{lhi} can be written in two equivalent representations
\begin{equation*}
L_t = \pobr{A^\infty_{\me k-r},L}_r = -\pobr{A^\infty_{<
k-r},L}_r.
\end{equation*}
So, we have to examine expansions of this hierarchy near $\infty$
and $v$ as well as at $0$, since the factor $p^{r}$ occur in
Poisson bracket. It turns out that first representation yields
direct access to terms with lowest orders, whereas the second
representation yields information about terms with highest orders.
Near $\infty$ we have
\begin{align*}
&L_t = (u_N)_t p^N + (u_{N-1}-N v)_t p^{N-1} + lower\ terms,\\
&L_t =-\pobr{A^\infty_{< k-r},L}_r = -\pobr{\alpha
p^{k-r-1}+l.t.,u_N p^N + l.t.}_r = (...) p^\alpha + l.t.\ ,
\end{align*}
where $\alpha = N+k-2$ for $N\neq 0$ when $r=k-1$; and $\alpha =
0$ for $N=0$ and $r=k-1$. This impose the constraints on fields
$u_N$ and $u_{N-1}$ given in Proposition. The expansion of
$A^\infty_{\me k-r}$ near $v$, is of the form $A^\infty_{\me k-r}
= higher\ terms + \gamma_1 (p-v) + \gamma_0$ as $A^\infty_{\me
k-r}$ does not have singularity at $v$. So, near $v$ we have
\begin{align*}
&L_t = higher\ terms + (u_{-m}+(m-1)v)_t (p-v)^{-m} + m v_t
(p-v)^{-m-1},\\
&L_t = \pobr{A^\infty_{\me k-r},L}_r = \pobr{h.t.+ \gamma_0 , h.t.
+ u_{-m}(p-v)^{-m}}_r = h.t. + (...)(p-v)^{-m-1}
\end{align*}
and the lowest order of left- and right-hand side of \eqref{lhi}
are always the same. The expansion of  $A^\infty_{\me k-r}$ near
$0$, is of the form $A^\infty_{\me k-r} = higher\ terms + \gamma
p^{k-r}$. So we have
\begin{align*}
&L_t = higher\ terms + \tfrac{1}{2}
\bra{\kr{\tfrac{d^2}{dp^2}L}{p=0}}_t p^2
+ \bra{\kr{\tfrac{d}{dp}L}{p=0}}_t p + \bra{\kr{L}{p=0}}_t,\\
&L_t = \pobr{A^\infty_{\me k-r},L}_r = \pobr{h.t.+ \gamma p^{k-r}
, h.t. + \kr{\tfrac{d}{dp}L}{p=0} p + \kr{L}{p=0}}_r = h.t. +
(...)p^\alpha.
\end{align*}
For $k=r=0$ we have $\alpha=0$ and there is no need of additional
constraints. For $k=1$ if $r\neq 1$: $\alpha = 0$ and both sides
have the same order in expansion at $0$. But for $k=r=1$ we have
$\alpha =1$. Hence, $(L|_{p=0})_t=0$ and we have to impose the
constraint $L|_{p=0}=0$. Then, both sides have the same form. For
$k=2$ and arbitrary $r$: $\alpha >0$ and the first constraint of
the form $L|_{p=0}=0$ is needed. Taking into consideration this
constraint: $\alpha=2$ and it follows that
$(\tfrac{d}{dp}L|_{p=0})_t=0$. Hence, both sides will agree if we
impose an additional constraint $\tfrac{d}{dp}L|_{p=0} =1$. For
$k=3$ the reasoning is similar to the case $k=2$, but there will
be one more constraint of the form $\tfrac{d^2}{dp^2}L|_{p=0}=0$
needed. Commutation of Lax hierarchies follows from
Corollary~\ref{cor}.
\end{proof}
\begin{proposition}
The case $k=r=0$ of Proposition \ref{apolv} by the transformation
$p\map p-v$ turns to the case $r=0$,~$k=1$ of Proposition
\ref{apol0}. Thus both Lax hierarchies are equivalent.
\end{proposition}
\begin{proof}
Consider the transformation $p'=p-v$, $x'=x$, $t'=t$, where
$t=t_q$~or~$\xi_q$. Then, $\pr_{p} = \pr_{p'}$, $\pr_x = \pr_{x'}-
v_x \pr_{p'}$ and $\pr_t = \pr_{t'} - v_t \pr_{p'}$. The points at
$\infty$ and $v$ transform into points at $\infty$ and $0$,
respectively and the Poisson bracket \eqref{pb} for $r=0$ is
preserved:
\begin{equation*}
\pobr{\cdot, \cdot}_0 = \pr_p \wedge \pr_x = \pr_{p'} \wedge
(\pr_{x'} + v_{x'} \pr_{p'}) = \pr_{p'} \wedge \pr_{x'} =
\pobr{\cdot, \cdot}_0' .
\end{equation*}
Let $L$ be the Lax function of the form \eqref{polv} from
Proposition \ref{apolv} for $r=k=0$. Then, by the above
transformation $L'=L$ is a Lax function of the form \ref{pol0}
from Proposition \ref{apolv} for $r=0$,~$k=1$. For meromorphic
function $A\in \Alg$, let $\bra{A}^\lambda_0$ mean the zero-order
term of Laurent series at $\lambda$. From \eqref{lhi} and
\eqref{lhv} it follows that
\begin{equation*}
v_{t_q} = \bra{\bra{L^{\tfrac{q}{N}}}^\infty_0}_x\qquad v_{\xi_q}
= \bra{\bra{L^{\tfrac{q}{m}}}^v_0}_x,
\end{equation*}
respectively. Thus the left- and right-hand side of \eqref{lhv}
are equal
\begin{align*}
&L_{\xi_q} = {L'}_{\xi_q'} - v_{\xi_q} L'_{p'} = {L'}_{\xi_q'} -
\bra{\bra{L^{\tfrac{q}{m}}}^v_0}_x L'_{p'} = {L'}_{\xi_q'} +
\pobr{\bra{L'^{\tfrac{q}{m}}}^0_0,L'}_0',\\
&L_{\xi_q} = -\pobr{\bra{L^{\tfrac{q}{m}}}^v_{<0},L}_0 = -
\pobr{\bra{L'^{\tfrac{q}{m}}}^0_{< 0},L'}_0'.
\end{align*}
Hence,
\begin{equation*}
L'_{\xi_q} =  - \pobr{\bra{L'^{\tfrac{q}{m}}}^0_{< 1},L'}_0'.
\end{equation*}
Similar calculations are valid at $\infty$ .
\end{proof}

Notice that, for the case $k=r=0$ of Proposition \ref{apolv} one
is able to construct Lax hierarchies related to the roots of $L$,
which is not possible for the case $k=1,r=0$ of Proposition
\ref{apol0}. In the sense, the first case is more general.

\begin{proposition}\label{pp1}
Under the transformation $p'=p^{-1}$, the following equalities
between some cases from Proposition \ref{apolv} hold:
\begin{itemize}
\item the Lax hierarchy related to $0$ for $k=3,r=2$ is equivalent
to the Lax hierarchy related to $\infty$ for $k=r=0$ with $N=-1$;
\item the Lax hierarchy related to $0$ for $k=2,r\neq 1$ with
$N=0$ is equivalent to the Lax hierarchy related to $\infty$ for
$k=1,r\neq 1$ with $N=-1$; \item Lax hierarchies related to
$\infty$ and $0$ for $k=2,r=1$ with $N=-1$ are equivalent to Lax
hierarchies related to $0$ and $\infty$ for $k=1,r=1$ with $N=-1$,
$L|_{p=0}=0$, respectively.
\end{itemize}
\end{proposition}
\begin{proof}
The appropriate Lax function from Proposition \ref{apolv} for
$k=3$, $r=2$ has the form
\begin{equation*}
L = u_0 + u_{-1} (p-v)^{-1} + ... + u_{-m} (p-v)^{-m}
\end{equation*}
where $L|_{p=0}=0$, $\tfrac{d}{dp}L|_{p=0}=1$ and
$\tfrac{d^2}{dp^2}L|_{p=0}=0$. Taking into consideration the above
constraints, expansion of $L$ around $0$ is $L = ... + (...)p^2 +
p $. By transformation $p'=p^{-1}$ we have that
\begin{equation*}
(p-v)^{-1} = (p'^{-1}-v)^{-1} = -v'-v'^2(p'-v')^{-1}
\end{equation*}
where $v'=v^{-1}$. Thus $L$ transforms into
\begin{equation*}
L' = u'_0 + u'_{-1} (p'-v')^{-1} + ... + u'_{-m} (p'-v')^{-m}.
\end{equation*}
From the expansion around $0$ of $L$ it follows that expansion of
$L'$ near $\infty$ is $L' = p'^{-1} + (...)p'^{-2} + ...\ $.
Hence, $u'_0=0$, $u'_{-1}=1$ and the Lax function $L'$ is an
appropriate one for $k=r=0$. Analogously for two next relations in
the proposition. The rest holds by Proposition \ref{0i}.
\end{proof}

Now, let us pass to the Hamiltonian formulation of Lax hierarchies
related to the appropriate Lax functions from Proposition
\ref{apolv}. In general, the relevant cases are for $k=0,1,2$.
Further we will consider only them. The differential at $\infty$
of functional $H$ for the Lax function of the general form
\eqref{polv} is given by
\begin{equation}\label{dHv}
dH\equiv d_\infty H = p^r \bra{\frac{1}{mu_{-m}}\bra{\var{H}{v}+
\sum_{i=1-m}^Niu_i\var{H}{u_{i-1}}}(p-v)^m +
\sum_{i=1-m}^{N+1}\var{H}{u_{i-1}}(p-v)^{-i}}
\end{equation}
as
\begin{align*}
\Tr_\infty \bra{L_t dH} &= -\int_\Omega \res_\infty \bra{p^{-r}L_t dH} dx \overset{\text{by \eqref{rest}}}{=} \int_\Omega \res_v \bra{p^{-r}L_t dH} dx\\
 &= \int_\Omega \bra{\sum_{i=-m}^N(u_i)_t\var{H}{u_i} + v_t\var{H}{v}} dx.
\end{align*}
For the Lax functions with constraints from Proposition
\ref{apolv} one has to modify differentials \eqref{dHv} in an
appropriate way or construct them by \eqref{diff}, i.e. the same
as in the next subsection. One has to examine when a given Lax
function from Proposition span the proper subspace with respect to
Poisson tensors. The procedure is rather technical and similar to
the proof of this proposition. Thus, we omit it and we will
present only the final results. The Lax functions from Proposition
\ref{apolv} for $k=0,1,2$ span proper subspace w.r.t. linear
Poisson tensor $n=0$ if $N\me 2r-2k+1$, $m\me -1$ and $r\me k$.
Then, it is given by \eqref{lin}. If it is not the case, Dirac
reduction is required. The reduced linear Poisson tensor for
$N=-1$, $m\me 1$ and $k=r=0,1,2$ is given by \eqref{lin_II}. These
Lax functions do not form a proper subspace w.r.t. quadratic
Poisson tensor $n=1$ and always the Dirac reduction procedure is
needed. For $k=r=0,1,2$ reduced quadratic Poisson tensors have the
form \eqref{quad_I}.

\begin{example} Two-field system: $k=r=1$.

The Lax function, taking into consideration appropriate
constraints, is given by the form
\begin{equation*}
L = (p-v) + u + v(u-v)(p-v)^{-1} = \frac{p(p+u-2v)}{p-v}.
\end{equation*}
For $\infty$ one finds $(L)^\infty_{\me 0} = p+u-v$ and the
following equation
\begin{align*}
&L_{t_1} = \pobr{\bra{L}^\infty_{\me 0},L}_1
\Longleftrightarrow\\
&\pmatrx{u\\ v}_{t_1} = \pmatrx{2u_xv+uv_x-2vv_x\\ u_xv} = \pi_0
dH^\infty_1 = \pi^{red}_1 dH^\infty_0 .
\end{align*}
The Lax hierarchy related to $v$ is the same as $L =
(L)^\infty_{\me 0} + (L)^v_{< 0}$. The Lax function has two roots
$0$ and $2v-u$. Then, for $(L^{-1})^0_{< 0} =
\frac{v}{2v-u}p^{-1}$ we have
\begin{align*}
L_{\tau_{-1}} = -\pobr{\bra{L^{-1}}^0_{< 0},L}_1
\Longleftrightarrow
\pmatrx{u\\ v}_{\tau_1} = \pmatrx{\frac{v_x}{2v-u}\\
\frac{2vv_x-u_xv}{(u-2v)^2}} = \pi_0 dH^0_{-1} = \pi^{red}_1
dH^0_{-2}.
\end{align*}
The Lax hierarchy related to the root $2v-u$ is up to the sign the
same as above since $L^{-1} = (L^{-1})^0_{\me 0} +
(L^{-1})^{2v-u}_{< 0}$.

The general form for a differential of a given functional $H$
according to \eqref{dHv} is
\begin{align*}
dH =
\frac{(2-u)\var{H}{u}+v\var{H}{v}}{(u-v)v^2}p(p-v)+\frac{1}{v}\var{H}{v}p.
\end{align*}
The Lax function defines the proper subspace w.r.t. the linear
Poisson tensor \eqref{lin}. The reduced quadratic Poisson tensors
is given by \eqref{quad_II}. Then,
\begin{align*}
\pi_0 = \pmatrx{\pr v+v\pr & \pr v\\
v\pr & 0 }\qquad \pi^{red}_1 = \pmatrx{2\pr uv + 2uv\pr & u\pr v+2v\pr v\\
v\pr u +2v\pr v & 2v\pr v} .
\end{align*}
The respective Hamiltonians are
\begin{align*}
&H^\infty_0 = \int_\Omega (u-v)\ dx\qquad H^\infty_1 =
\frac{1}{2}\int_\Omega
(u^2-v^2)\ dx\\
&H^0_{-2} = \int_\Omega \frac{v-u}{(u-2v)^2}\ dx\qquad H^0_{-1} =
\int_\Omega \ln \bra{\frac{u}{v}-2}\ dx .
\end{align*}
\end{example}

\subsection{Rational Lax functions}

Let us consider the general form of meromorphic Lax function given
by
\begin{equation}\label{rat}
L = \sum_{k=-m_0}^{N}u_k p^k + \sum_{i=1}^\alpha
\sum_{k_i=1}^{m_i} a_{i,k_i}(p-v_i)^{-k_i}
\end{equation}
where $u_k$, $a_{i,k_i}$ and $v_i$ are dynamical fields. From this
class of functions considered in the following subsection we
exclude those which have been examined earlier, i.e. \eqref{pol0}
and \eqref{polv}. Any function \eqref{rat} in general has a pole
at $\infty$ of order $N$, at $0$ of order $m_0$ and $\alpha$
evolution poles at $v_j$ of order $m_j$. Then, one can construct
positive powers of Laurent series at poles of $L$. Negative powers
can be constructed as expansion at the roots of $L$. These powers
generate for appropriate $r$ and $k$ Lax hierarchies
\eqreff{lhi}{lhv}.
\begin{proposition}\label{arat}
Function of the form \eqref{rat} is an appropriate one in the
following cases:
\begin{enumerate}
    \item $k=r=0$:
\begin{itemize}
    \item $N\me 1$, $u_N=1$, $u_{N-1}=0$, $m_0=0$,
    \item $\forall_k\ u_k=0$, $\sum_{i=1}^\alpha a_{i,1}=1$,
    $\sum_{i=1}^\alpha \bra{a_{i,1}v_i + a_{i,2}}=0$;
\end{itemize}
    \item $k=1$, $r\in \bb{Z}$:
\begin{itemize}
    \item $N\me 1$, $u_N=1$, $m_0\me 1$,
    \item $N=-1$,  $u_{-1} + \sum_{i=1}^\alpha a_{i,1}=1$, $m_0\me 1$,
    \item $N\me 1$, $u_N=1$, $m_0=0$, $L|_{p=0}=0$ for $r=1$,
    \item $\forall_k\ u_k=0$, $\sum_{i=1}^\alpha a_{i,1}=1$, $L|_{p=0}=0$ for $r=1$;
\end{itemize}
\item $k=2$, $r\in \bb{Z}$:
\begin{itemize}
    \item  $N\me 1$, $m_0\me 1$, $u_{-m_0}=1$,
    \item  $N\me 1$, $m_0=0$, $L|_{p=0}=0$,
    $\tfrac{d}{dp}L|_{p=0}=1$,
    \item  $N=0$, $u_0=0$ for $r=1$, $m_0\me 1$, $u_{-m_0}=1$,
    \item  $\forall_{k\neq 0}\ u_k=0$, $u_0=0$ for $r=1$,
    $L|_{p=0}=0$, $\tfrac{d}{dp}L|_{p=0}=1$;
\end{itemize}
    \item  $k=3$ and $r=2$
\begin{itemize}
    \item $N=0$, $m_0\me 1$, $u_{1-m_0}=0$, $u_{-m_0}=1$,
    \item $\forall_{k\neq 0}\ u_k=0$, $\kr{L}{p=0}=0$,
    $\kr{\tfrac{d}{dp}L}{p=0}=1$, $\kr{\tfrac{d^2}{dp^2}L}{p=0}=0$.
\end{itemize}
\end{enumerate}
Moreover Lax hierarchies calculated at different points for the
same $r$ and $k$ mutually commute. We excluded here the Lax
functions of the form \eqref{pol0} and \eqref{polv}.
\end{proposition}
The proof is similar to the one of Proposition \ref{apolv}. The
evolution of the general form of rational function \eqref{rat} is
given by
\begin{equation*}
L_t = \sum_{k=-m_0}^{N}(u_k)_t p^k + \sum_{i=1}^\alpha
\sum_{k_i=1}^{m_i}\bra{ (a_{i,k_i})_t + (k_i - 1) a_{i,k_i-1}
(v_i)_t} (p-v_i)^{-k_i} + \sum_{i=1}^\alpha m_i a_{i,m_i} (v_i)_t
(p-v_i)^{-m_i-1} .
\end{equation*}
The Lax hierarchies have to be examined near $\infty$, $0$ and
dynamical poles. So, the meromorphic function of the form
\eqref{rat} will be an appropriate Lax function if:
\begin{itemize}
    \item the right-hand sides of the Lax hierarchy considered
    and the time derivatives $L_t$ will have the same
order at all above poles,
    \item the number of independent equations, resulting from Lax hierarchies,
    will be the same as that of dynamical coefficients included in $L$.
\end{itemize}
The first condition implies all constraints considered in
Proposition. To see that, an analysis like in proof of Proposition
\ref{apolv} is needed. So, by the first condition the right-hand
sides of considered Lax hierarchies for appropriate $r$ and $k$
can be uniquely presented in the form of $L_t$, i.e. the left-hand
sides. So, the second condition immediately follows from the first
one.

The simplest way of deriving dispersionless systems related to a
given meromorphic Lax functions is to transform Lax hierarchies
into purely polynomial form in $p$ through removal of finite
singularities. It can be done by multiplication of both sides of
Lax hierarchies by a proper factor.

\begin{proposition}\label{pp}
Under transformation $p'=p^{-1}$ Lax hierarchies, from Proposition
\ref{arat}, defined at $\infty$ and $0$ for appropriate $r$ and
$k$ transforms into Lax hierarchies defined at $0$ and $\infty$
for $r'=2-r$ and $k'=3-k$, respectively.
\end{proposition}
See proof of Proposition \ref{pp1} and the comment after
Proposition \ref{pp2}.

Hence, the relevant cases from Proposition \ref{arat} are:
\begin{itemize}
    \item $k=0$, $r=0$;
    \item $k=1$, $r\in \bb{Z}\backslash \{0\}$;
    \item $k=2$, $r=2$.
\end{itemize}
The remaining cases can be obtained from the above cases by the
transformation $p'=p^{-1}$ according to Proposition \ref{pp}.

Once again we will consider Poisson tensors defined at $\infty$.
This time we are not going to present the explicit form of
differentials $d_\infty H$ for the general meromorphic Lax
function, but we will explain how to construct them. We postulate
that
\begin{equation}\label{diff}
  dH\equiv d_\infty H = \sum_{i=N_\infty-\beta+1}^{N_\infty} \gamma_i\ p^{r-i-1}
\end{equation}
where $\beta$ is a number of dynamical coefficients in $L$ and
$N_\infty$ is the highest order of Laurent series of $L_t$ at
$\infty$. The form \eqref{diff} allows us to solve \eqref{eucl}
($\lambda=\infty$) to obtain  functions $\gamma_i$ in terms of
dynamical coefficients of $L$ and its variational derivatives such
that we obtain the required Euclidean form. We will consider only
relevant cases of meromorphic Lax functions from Proposition
\ref{arat}. Verification that they span the proper subspace with
respect to Poisson tensors is similar to the proof of this
proposition. These Lax functions span the proper subspace w.r.t.
the linear Poisson tensor \eqref{lin} for $k=0$ if $N\me 1$ and
for $k=1,2$ if $N\me 2r-2k+1\me -m_0$. If not the case, the Dirac
reduction is required. The reduced linear tensors for $k=r=0,1,2$
and $N=-1$ ($N$ is the highest order of Laurent series of $L$ at
$\infty$) are given by
\begin{align}\label{lin_II}
\pi^0_{\infty}dH  &= \pobr{\bra{dH}^\infty_{\me 0},L}_r -
\bra{\pobr{dH,L}_r}^\infty_{\me r} +
\pobr{\gamma_1p+\gamma_0,L}_r\\
\notag &\gamma_1 = \pr_x^{-1}\res_\infty \pobr{dH,L}_0\\
\notag &\gamma_0 = \pr_x^{-1}\res_\infty \pobr{dH,L}_1 -
\pr_x^{-1}\bra{\gamma_1\bra{(L)_{-2}^\infty}_x+2(\gamma_1)_x(L)_{-2}^\infty},
\end{align}
where $(L)_{-2}^\infty$ is the coefficient staying at the term of
order $-2$ in Laurent series at $\infty$. For $k=r=0$ we have
$(L)_{-2}^\infty=0$ and \eqref{lin_II} simplified. Notice, that
for $k=r=0$ the reduced Poisson tensor \eqref{lin_II} is always
local, but for the remaining cases it is in general not. In the
case of quadratic Poisson tensors, considered Lax functions do not
span proper subspaces and the Dirac reduction is needed. The
reduced quadratic Poisson tensor for $k=r=0,1,2$ are given by
\eqref{quad_I}, and for $k=1, r=0$ and $k=2, r=1$ by
\eqref{quad_II} where $m=m_0$.

\begin{example} The two-field system: $k=r=0$.

Let the Lax function have the form
\begin{align*}
L = u(p-v)^{-1} + (1-u)\bra{p-\frac{uv}{u-1}}^{-1}.
\end{align*}
The roots of $L$ are $\infty$ and $\alpha = \frac{2uv-v}{u-1}$.
Then, for $\infty$ we have
\begin{align*}
L_{t_2} = \pobr{\bra{L^{-2}}^\infty_{\me 0},L}_0
\Longleftrightarrow \pmatrx{u\\ v}_{t_2} = \pmatrx{2uv\\
\frac{(3u-1)v^2}{u-1}}_x = \pi^{red}_0 dH^\infty_2 = \pi^{red}_1
dH^\infty_1,
\end{align*}
where $(L^{-2})^\infty_{\me 0} = p^2+\frac{2uv^2}{u-1}$. The
hierarchy for $\alpha$ is the same as $(L^q)^\infty_{\me 0} = L -
(L^q)^\alpha_{< 0}$. The function $L$ has poles at $v$ and
$\lambda = \frac{uv}{u-1}$. At the point $v$ one finds the
following system
\begin{align*}
L_{\xi_1} = -\pobr{\bra{L}^v_{< 0},L}_0 \Longleftrightarrow
\pmatrx{u\\ v}_{\xi_1} = \pmatrx{\frac{u(u-1)^3}{v^2}\\
\frac{(u-1)^2}{v}}_x = \pi^{red}_0 dH^v_1 = \pi^{red}_1 dH^v_0,
\end{align*}
where $(L)^v_{< 0} = u(p-v)^{-1}$. The Lax function is invariant
with respect to the transformation $u\map 1-u$, $v\map
\frac{uv}{u-1}$. Therefore, the Lax hierarchy related to $\lambda$
can be obtained through this transformation.

In the case the differential of a given functional calculated by
\eqref{diff} is
\begin{align*}
dH =
\bra{\frac{2(u-1)^3}{v^3}\var{H}{u}+\frac{(u-1)^2}{uv^2}\var{H}{v}}p^3
+
\bra{\frac{3(u-1)^2(2u-1)}{v^2}\var{H}{u}+\frac{(u-1)(3u-1)}{uv}\var{H}{v}}p^2
.
\end{align*}
Then, from \eqref{lin_II} and \eqref{quad_I} we find the following
Poisson tensors
\begin{align*}
\pi^{red}_0 = \pmatrx{0 & \pr (1-u)\\ (1-u)\pr & -\pr
v-v\pr}\qquad \pi^{red}_1 = \pmatrx{\pr
\frac{u}{v^2}(u-1)^3+\frac{u}{v^2}(u-1)^3\pr &
\frac{(u-1)^2}{v}\pr\\ \pr \frac{(u-1)^2}{v} & 0},
\end{align*}
respectively. The Hamiltonians are
\begin{align*}
H^\infty_1 = \int_\Omega \frac{uv}{1-u}\ dx\qquad H^\infty_2 =
\int_\Omega \frac{uv^2}{1-u}\ dx\qquad H^v_0 = \int_\Omega v\
dx\qquad H^v_1 = \int_\Omega \frac{u(u-1)^2}{v}\ dx .
\end{align*}
\end{example}

\begin{example} The four-field dispersionless system: $k=r=0$.

For the Lax function of the form
\begin{align*}
L = p + a(p-v)^{-1} + b(p-w)^{-1}
\end{align*}
near $\infty$ one finds
\begin{align*}
L_{t_2} = \pobr{\bra{L^{2}}^\infty_{\me 0},L}_0
\Longleftrightarrow
\pmatrx{a\\ b\\ v\\ w}_{t_2} = \pmatrx{2av\\ 2bw\\ 2a+2b+v^2\\
2a+2b+w^2}_x = \pi_0 dH^\infty_2 = \pi^{red}_1 dH^\infty_1,
\end{align*}
where $(L^2)^\infty_{\me 0} = p^2 +2a+2b$. Near to the $v$ we have
\begin{align*}
L_{\xi_1} = -\pobr{\bra{L}^v_{< 0},L}_0 \Longleftrightarrow
\pmatrx{a\\ b\\ v\\ w}_{t_1} =
 \pmatrx{a - \frac{ab}{(v-w)^2}\\ \frac{ab}{(v-w)^2}\\
 v + \frac{b}{v-w}\\ \frac{a}{v-w}}_x = \pi_0 dH^v_1 = \pi^{red}_1 dH^v_0,
\end{align*}
where $(L)^v_{\les 0} = a(p-v)^{-1}$. There are three, very
complicated, roots of $L$. Thus, we are not going to calculate the
respective equations.

The differential of a functional $H$ is
\begin{align*}
dH = &
\bra{\frac{2}{v-w}\var{H}{b}+\frac{1}{b}\var{H}{w}}\frac{(p-v)^2(p-w)}{(v-w)^2}
+\bra{\frac{2}{w-v}\var{H}{a}+\frac{1}{a}\var{H}{v}}\frac{(p-v)(p-w)^2}{(w-v)^2}\\
&+\frac{1}{(v-w)^2}\var{H}{b}(p-v)^2+\frac{1}{(w-v)^2}\var{H}{a}(p-w)^2.
\end{align*}
Then, from \eqref{lin} and \eqref{quad_I} one finds the linear
\begin{align*}
\pi_0 = \pmatrx{0 & 0 & \pr & 0\\
0 & 0 & 0 & \pr\\ \pr & 0 & 0 & 0\\
0 & \pr & 0 & 0}
\end{align*}
and quadratic Poisson tensors
\begin{align*}
\pi^{red}_1 = \pmatrx{\pr a+a\pr -\pr
\frac{ab}{(v-w)^2}-\frac{ab}{(v-w)^2}\pr & \pr
\frac{ab}{(v-w)^2}+\frac{ab}{(v-w)^2}\pr & (v+\frac{b}{v-w})\pr &
\frac{a}{v-w}\pr\\
\pr \frac{ab}{(v-w)^2}+\frac{ab}{(v-w)^2}\pr & \pr b+b\pr -\pr
\frac{ab}{(v-w)^2}-\frac{ab}{(v-w)^2}\pr & -\frac{b}{v-w}\pr &
(w-\frac{a}{v-w})\pr\\
\pr(v+\frac{b}{v-w}) & -\pr \frac{b}{v-w} & 2\pr & \pr\\
\pr \frac{a}{v-w}& \pr (w-\frac{a}{v-w}) & \pr & 2\pr},
\end{align*}
respectively. The Hamiltonians are
\begin{align*}
&H^\infty_1 = \int_\Omega (av+bw)\ dx\qquad H^\infty_2 =
\int_\Omega \bra{(a+b)^2+av^2+bw^2}\ dx\\
&H^v_0 = \int_\Omega a\ dx\qquad H^v_1 = \int_\Omega
\bra{av+\frac{ab}{v-w}}\ dx .
\end{align*}
\end{example}

\begin{example} Four-field system: $k=1$, $r\in \bb{Z}$.
\label{ex}

Let us consider a Lax function of the form
\begin{equation}\label{a1}
L= p+u+vp^{-1}+w(p-s)^{-1}.
\end{equation}
It has poles at $\infty$, $0$ and $w$. Related equations to
$\infty$ for $(L^{2-r})^\infty_{\me 1-r} = p^{2-r}+(2-r)up^{1-r}$
are
\begin{align*}
&L_{t_{2-r}} = \pobr{\bra{L^{2-r}}^\infty_{\me 1-r},L}_r \Longleftrightarrow\\
&\pmatrx{u\\ v\\ w\\ s}_{t_{2-r}} = (2-r)\pmatrx{(1-r)uu_x+v_x+w_x\\ u_xv+(1-r)uv_x\\
u_xw+(1-r)uw_x+(ws)_x\\ u_xs+(1-r)us_x+ss_x} =
\pi_0dH^\infty_{2-r} = \pi^{red}_1 dH^\infty_{1-r} .
\end{align*}
The first equations from Lax hierarchies related to $0$ for $r\neq
0$ are
\begin{align*}
L_{\tau_{r}} = -\pobr{\bra{L^{r}}^0_{< 1-r},L}_r
\Longleftrightarrow
\pmatrx{u\\ v\\ w\\ s}_{\tau_{r}} =  rv^r \pmatrx{\ln v\\ u-\frac{w}{s}\\
\frac{w}{s}\\ \ln \frac{s}{v}}_x = \pi_0dH^0_{r} =
\pi^{red}_1dH^0_{r-1},
\end{align*}
where $(L^r)^0_{\les 1-r} = v^r p^{-r}$. But for $r=0$ we have
\begin{align*}
L_{\tau_1} = -\pobr{\bra{L}^0_{< 1},L}_0 \Longleftrightarrow
\pmatrx{u\\ v\\ w\\ s}_{\tau_1} =  \pmatrx{u-\frac{w}{s}\\ v-\frac{vw}{s^2}\\
\frac{vw}{s^2}\\ \frac{w-v}{s}-u}_x = \pi_0dH^0_1 =
\pi^{red}_1dH^0_0,
\end{align*}
where $(L)^0_{\les 1} = u-\frac{w}{s}+ v p^{-1}$. For $r=1$ and
$(L)^s_{\les 0} = w(p-s)^{-1}$ one finds
\begin{align*}
L_{\xi_1} = -\pobr{\bra{L}^s_{< 0},L}_1 \Longleftrightarrow
\pmatrx{u\\ v\\ w\\ s}_{\xi_1} =  \pmatrx{w_x\\ v\bra{\frac{w}{s}}_x\\
u_xw+(ws)_x-v\bra{\frac{w}{s}}_x\\ u_xs+ss_x+s\bra{\frac{v}{s}}_x}
= \pi_0dH^s_1 = \pi^{red}_1dH^s_0 .
\end{align*}
Once again we are not going to consider Lax hierarchies related to
roots of $L$.

The differential of a functional $H$ is
\begin{align*}
dH = \bra{\var{H}{s} +
\frac{1}{s^2}\bra{\var{H}{v}-\var{H}{w}}}p^{r+2} -
\bra{\frac{2}{s}\bra{\var{H}{v}-\var{H}{w}}+\frac{1}{w}\var{H}{s}}p^{r+1}
+\var{H}{v}p^r + \var{H}{u}p^{r-1}.
\end{align*}
The Lax function \eqref{a1} span the proper subspace w.r.t linear
Poisson tensor \eqref{lin} only for $r=0,1$. The reduced quadratic
tensors are for $r=0,1$ given by \eqref{quad_II} and
\eqref{quad_I}, respectively. Thus, for $r=0$:
\begin{equation}\label{p1}
\pi_0 = \pmatrx{0 & \pr & 0 & 0\\ \pr & 0 & 0 & -\pr\\ 0 & 0 & 0 &
\pr\\ 0 & -\pr & \pr & 0}
\end{equation}
and
\begin{equation}\label{p2}
\pi^{red}_1 = \pmatrx{2\pr & \pr(u-\frac{w}{s}) & \pr \frac{w}{s} & -\pr\\
(u-\frac{w}{s})\pr & \pr v(1-\frac{w}{s^2})-v(1-\frac{w}{s^2})\pr
& \pr \frac{vw}{s^2}+\frac{vw}{s^2}\pr & (\frac{w-v}{s}-u)\pr\\
\frac{w}{s}\pr & \pr \frac{vw}{s^2}+\frac{vw}{s^2}\pr
& \pr w(1-\frac{v}{s^2})-w(1-\frac{v}{s^2})\pr & (u+s+\frac{v-w}{s})\pr\\
-\pr & \pr (\frac{w-v}{s}-u) & \pr (u+s+\frac{v-w}{s}) & 2\pr} .
\end{equation}
The related Hamiltonians are
\begin{align*}
&H^\infty_1 = \int_\Omega \bra{uv+uw+ws}\ dx\qquad H^\infty_2 =
\int_\Omega \bra{u^2v+v^2+ws^2+2uws+u^2w+2vw+w^2}\
dx\\
&H^0_0 = \int_\Omega v\ dx\qquad H^0_1 = \int_\Omega
v\bra{u-\frac{w}{s}}\ dx .
\end{align*}
For $r=1$ we have:
\begin{equation*}
\pi_0 = \pmatrx{0 & \pr v & \pr w & \pr s\\ v\pr & 0 & 0 & 0\\
w\pr & 0 & s\pr w+w\pr s & s\pr s\\ s\pr & 0 & s\pr s & 0}
\end{equation*}
and
\begin{equation*}
\pi^{red}_1 = \pmatrx{\pr(v+w)+(v+w)\pr & u\pr v & 2\pr ws+ws\pr +u\pr w & (u+s)\pr s\\
v\pr u & 2v\pr v & 2v\pr w & v\pr s\\
\pr ws+2ws\pr +w\pr u & 2w\pr v & \pi_{ww} & (s^2+us+v+2w)\pr s\\
s\pr (u+s) & s\pr v & s\pr (s^2+us+v+2w) & 2s\pr s},
\end{equation*}
where $\pi_{ww} = \pr uws+uws\pr+w\pr (2s^2+w)+(2s^2+w)\pr w$. The
related Hamiltonians are
\begin{align*}
&H^\infty_0 = \int_\Omega u\ dx\qquad H^\infty_1 = \int_\Omega
\bra{\frac{1}{2}u^2+v+w}\ dx\\
&H^0_0 = \int_\Omega (u-\frac{w}{s})\ dx\qquad H^0_1 = \int_\Omega
\bra{\frac{1}{2}u^2+v-\frac{uw}{s}-\frac{vw}{s^2}+\frac{w^2}{2s^2}}\
dx\\
&H^s_0 = \int_\Omega \frac{w}{s}\ dx\qquad H^s_1 = \int_\Omega
\bra{w+\frac{uw}{s}+\frac{vw}{s^2}-\frac{w^2}{2s^2}}\ dx.
\end{align*}
\end{example}

\pagebreak

\begin{example} Four-field system: $k=r=2$.

Lax function \eqref{a1} transformed by $p\map p^{-1}$ has the form
\begin{equation*}
L= vp+u-\frac{w}{s}+p^{-1}-\frac{w}{s^2}(p-s^{-1})^{-1}.
\end{equation*}
For $(L)^{s^{-1}}_{\les 0} = \frac{w}{s^2}(p-s^{-1})^{-1}$ one
finds the system
\begin{align*}
L_{\xi_1} = -\pobr{\bra{L}^{s^{-1}}_{< 0},L}_1 \Longleftrightarrow
\pmatrx{u\\ v\\ w\\ s}_{\xi_1} =  \pmatrx{ -\frac{w}{s}\\ -\frac{vw}{s^2}\\
w\bra{\frac{v}{s^2}-1}\\ -u-s+\frac{v-w}{s}}_x =
\pi_0dH^{s^{-1}}_1 = \pi^{red}_1dH^{s^{-1}}_0
\end{align*}
commuting, by Proposition \eqref{pp}, with equations from Example
\ref{ex} for $r=0$. The linear and quadratic Poisson tensors are
given by \eqref{p1} and \eqref{p2}, respectively. Hamiltonian
functionals are given by
\begin{align*}
H^{s^{-1}}_0 = -\int_\Omega w\ dx\qquad H^{s^{-1}}_1 = \int_\Omega
w\bra{u+\frac{v-1}{s}}\ dx.
\end{align*}
\end{example}

\section{Comments}

In the article we have presented a systematic construction of
multi-Hamiltonian dispersionless systems with meromorphic Lax
representations. It is shown that for a given meromorphic Lax
function $L$, if allowed by $k$ and $r$, one can construct Lax
hierarchies related to all poles of $L$ and $L^{-1}$. These Lax
hierarchies, if we fix $k$ and $r$, mutually commute. It is shown
how to construct Poisson tensors and infinite hierarchies of
constants of motion. It is proved that Poisson tensors, from the
original function space, reconstructed for different poles are
equal. Also, we have examined systematically the forms of
appropriate meromorphic Lax functions, with finite number of
dynamical fields, allowing construction of consistent
dispersionless systems. The Poisson tensors constructed for the
appropriate meromorphic Lax functions considered in the following
article are nondegenerate.

Articles \cite{FS}-\cite{S} deal with rational Lax functions from
the algebra with fixed Poisson bracket $r=1$. However, only the
Lax hierarchies generated by powers constructed near $\infty$ have
been considered there. For the class of rational Lax functions
used in these papers the bi-Hamiltonian structures are degenerate,
i.e. the determinants of the related metrics vanish. The reason is
that the constraint of the form $L|_{p=0} = 0$ is not taken into
consideration. So, one dynamical field always can be represented
as a function of all others. This fact entails the degeneracy of
Poisson tensors.

There is a different approach to meromorphic Lax functions. From
the complex analysis it is well known that meromorphic function
can be uniquely presented in the factorized form. Because of such
a factorization there is no problem in finding poles of $L$ and
$L^{-1}$ near which one construct powers and related Lax
hierarchies. Another advantage is that the dispersionless systems
obtained have very symmetrical form. However, the disadvantage is
that Poisson tensors are significantly more complicated. Such,
factorized form of Lax functions as well allows for finding new
reductions which are not obvious when we have Lax function in the
standard form, see \cite{FS}-\cite{S}.

In the paper we have considered dispersionless systems with a
finite number of dynamical fields. However, Lax function being
infinite formal Laurent series leads to the construction of
dispersionless infinite-field Benney moment like equations. Such
systems for Laurent series at $\infty$ have been considered
earlier in \cite{Bl2}. The original Benney moment equation can be
obtained for $k=r=0$. If we consider formal Laurent series at a
pole being a dynamical field $v(x)$ we will construct new classes
of infinite-field dispersionless systems. They, together with
bi-Hamiltonian structures, will be studied in a forthcoming
article. Furthermore, all finite-field dispersionless systems,
with meromorphic Lax functions, considered in this paper may be
considered as reductions of these infinite-field systems.

All Lax functions used in the article belong to the algebras of
meromorphic functions. But, it is straightforward to extend the
theory presented into algebras of holomorphic functions. So, it
may be worth looking systematically for new classes of appropriate
Lax functions being holomorphic and allowing construction of
related dispersionless systems.

Another issue is the extension of the theory of meromorphic Lax
representations presented for dispersionless systems in order to
construct integrable dispersive soliton systems for rational Lax
operators. The first approach towards this was made in article
\cite{EOR}. However, the authors constructed soliton systems
related only to the case $k=r=0$ from our article. A more general
theory, of dispersive deformations of formal Lax functions being
polynomials in $p$ and $p^{-1}$, is presented in our paper
\cite{BS}. This approach is based on the Weyl-Moyal-like
quantization procedure. The idea relies on the deformation of the
usual multiplication in the algebra $\Alg$ to the new associative
but non-commutative $\star$-product. However, this theory works
only for $r=0,1,2$. Deformations of Poisson algebras for $r=0,2$
are equivalent and lead to the construction of field soliton
systems, but for $r=1$ they lead to the construction of lattice
soliton  systems. So, in a forthcoming article we are going to
present a general theory of the field and lattice soliton systems
for rational Lax operators.

\subsection*{Acknowledgment}

This work was partially supported by KBN research grant No. 1 P03B
111 27. One of the authors B. M. Sz. would like to thank M. Pavlov
and A. Yu. Orlov for some discussions and useful references.

\end{document}